\documentclass[longauth]{aa}  

\usepackage{graphicx}
\usepackage{txfonts}
\usepackage{hyperref}
\hypersetup{
    colorlinks=true,
    linkcolor=blue,
    filecolor=blue,      
    urlcolor=blue,
    citecolor=blue,
    pdftitle={Dark lens candidates from Gaia Data Release~3},
    pdfpagemode=FullScreen,
    }
\usepackage[normalem]{ulem}
\usepackage{multirow}
\usepackage{amsmath}
\usepackage{booktabs}
\usepackage{xcolor}
\usepackage{threeparttablex}

\def\gaia{\textit{Gaia}\xspace}
\newcommand\gdr[1]{\gaia~DR#1}

\newcommand\ugdr[1]{\gaia{}DR3-ULENS-#1}

\def\gmag{$G$\xspace}
\def\gbp{$G_{\rm BP}$\xspace}
\def\grp{$G_{\rm RP}$\xspace}

\providecommand{\Msun}{\ensuremath{\,{M}_{\odot}}\xspace}

\def\tE{t_{\rm E}\xspace}

\def\piEN{\pi_{\rm EN}\xspace}
\def\piEE{\pi_{\rm EE}\xspace}
\def\piE{\pi_{\rm E}\xspace}

\newcommand{\changed}[1]{{#1}}

\titlerunning{Dark lens candidates from \gaia Data Release~3} % max 60
\authorrunning{Kruszy\'nska et al.}

\begin{document}

   \title{Dark lens candidates from \gaia Data Release~3}

   \subtitle{}

\author{K.~Kruszy{\'n}ska\inst{\ref{oauw}, \ref{lco}}\fnmsep\thanks{\email{kkruszynska@lco.global}}
    \and
    {\L}.~Wyrzykowski\inst{\ref{oauw}}
    \and
    K.~A.~Rybicki\inst{\ref{wis}}
    \and
    K.~Howil\inst{\ref{oauw}}
    \and
    M.~Jab{\l}o{\'n}ska\inst{\ref{oauw}, \ref{anu}}
    \and
    Z.~Kaczmarek\inst{\ref{heidelberg}}
    \and 
    N.~Ihanec\inst{\ref{oauw}}
    \and
    M.~Maskoli\={u}nas\inst{\ref{vilnius}}
    \and
    M.~Bronikowski\inst{\ref{cacung}}
    \and 
    U.~Pylypenko\inst{\ref{oauw}}
    \and
    A.~Udalski\inst{\ref{oauw}}
    \and
    P.~Mróz\inst{\ref{oauw}}
    \and
    R.~Poleski\inst{\ref{oauw}}
    \and
    J.~Skowron\inst{\ref{oauw}}
    \and
    M.~K.~Szyma{\'n}ski\inst{\ref{oauw}}
    \and
    I.~Soszy{\'n}ski\inst{\ref{oauw}}
    \and
    P.~Pietrukowicz\inst{\ref{oauw}}
    \and
    S.~Koz{\l}owski\inst{\ref{oauw}}
    \and
    K.~Ulaczyk\inst{\ref{warwick}}
    \and
    P.~Iwanek\inst{\ref{oauw}}
    \and
    M.~Wrona\inst{\ref{oauw}}
    \and
    M.~Gromadzki\inst{\ref{oauw}}
    \and
    M.~J.~Mr{\'o}z\inst{\ref{oauw}}
    \and
    F.~Abe\inst{\ref{nagoya}}
    \and
    K.~Bando\inst{\ref{oosaka}}
    \and 
    R.~Barry\inst{\ref{goddard}}
    \and
    D.~P.~Bennett\inst{\ref{goddard}, \ref{maryland}}
    \and
    A.~Bhattacharya\inst{\ref{goddard}, \ref{maryland}}
    \and
    I.~A.~Bond\inst{\ref{massey}}
    \and
    A.~Fukui\inst{\ref{depstoudai}}
    \and
    R.~Hamada\inst{\ref{oosaka}}
    \and
    S.~Hamada\inst{\ref{oosaka}}
    \and
    N.~Hamasaki\inst{\ref{oosaka}}
    \and
    Y.~Hirao\inst{\ref{ioatoudai}}
    \and
    S.~Ishitani~Silva\inst{\ref{goddard}, \ref{oak}}
    \and
    Y.~Itow\inst{\ref{nagoya}}
    \and
    N.~Koshimoto\inst{\ref{oosaka}}
    \and
    Y.~Matsubara\inst{\ref{nagoya}}
    \and
    S.~Miyazaki\inst{\ref{jaxa}}
    \and
    Y.~Muraki\inst{\ref{nagoya}}
    \and
    T.~Nagai\inst{\ref{oosaka}}
    \and
    K.~Nunota\inst{\ref{oosaka}}
    \and
    G.~Olmschenk\inst{\ref{goddard}}
    \and
    C.~Ranc\inst{\ref{iap}}
    \and
    N.~J.~Rattenbury\inst{\ref{auckland}}
    \and
    Y.~Satoh\inst{\ref{oosaka}}
    \and
    T.~Sumi\inst{\ref{oosaka}}
    \and
    D.~Suzuki\inst{\ref{oosaka}}
    \and
    P.~J.~Tristram\inst{\ref{mtjohn}}
    \and
    A.~Vandorou\inst{\ref{goddard}, \ref{maryland}}
    \and
    H.~Yama\inst{\ref{oosaka}}
    }

\institute{Astronomical Observatory, University of Warsaw, Al. Ujazdowskie 4, 00-478 Warszawa, Poland
  \label{oauw}
  \and
  Las Cumbres Observatory, 6740 Cortona Drive, Suite 102, Goleta, CA 93117, USA
  \label{lco}
  \and
  Department of Particle Physics and Astrophysics, Weizmann Institute of Science, Rehovot 76100, Israel
  \label{wis}
  \and
  Research School of Astronomy and Astrophysics, Australian National University, Mount Stromlo Observatory, Cotter Road Weston Creek, ACT 2611, Australia
  \label{anu}
  \and
  Zentrum f{\"u}r Astronomie der Universit{\"a}t Heidelberg, Astronomisches Rechen-Institut, M{\"o}nchhofstr. 12-14, 69120 Heidelberg, Germany
  \label{heidelberg}
  \and
  Institute of Theoretical Physics and Astronomy, Vilnius University, Saulėtekio al. 3, Vilnius, LT-10257, Lithuania
  \label{vilnius}
  \and
  Center for Astrophysics and Cosmology, University of Nova Gorica, Vipavska 11c, SI-5270 Ajdovščina, Slovenia
  \label{cacung}
  \and
  Department of Physics, University of Warwick, Gibbet Hill Road, Coventry, CV4~7AL,~UK 
  \label{warwick}
  \and
  Institute for Space-Earth Environmental Research, Nagoya University, Nagoya 464-8601, Japan
  \label{nagoya}
  \and
  Department of Earth and Space Science, Graduate School of Science, Osaka University, Toyonaka, Osaka 560-0043, Japan
  \label{oosaka}
  \and
  Code 667, NASA Goddard Space Flight Center, Greenbelt, MD 20771, USA
  \label{goddard}
  \and
  Department of Astronomy, University of Maryland, College Park, MD 20742, USA
  \label{maryland}
  \and
  Institute of Natural and Mathematical Sciences, Massey University, Auckland 0745, New Zealand
  \label{massey}
  \and
  Department of Earth and Planetary Science, Graduate School of Science, The University of Tokyo, 7-3-1 Hongo, Bunkyo-ku, Tokyo 113-0033, Japan
  \label{depstoudai}
  \and
  Institute of Astronomy, Graduate School of Science, The University of Tokyo, 2-21-1 Osawa, Mitaka, Tokyo 181-0015, Japan
  \label{ioatoudai}
  \and
  Oak Ridge Associated Universities, Oak Ridge, TN 37830, USA
  \label{oak}
  \and
  Institute of Space and Astronautical Science, Japan Aerospace Exploration Agency, 3-1-1 Yoshinodai, Chuo, Sagamihara, Kanagawa 252-5210, Japan
  \label{jaxa}
  \and
  Sorbonne Universit{\`e}, CNRS, UMR 7095, Institut d'Astrophysique de Paris, 98 bis bd Arago, 75014 Paris, France
  \label{iap}
  \and
  Department of Physics, University of Auckland, Private Bag 92019, Auckland, New Zealand
  \label{auckland}
  \and
  University of Canterbury Mt. John Observatory, P.O. Box 56, Lake Tekapo 8770, New Zealand
  \label{mtjohn}
 }

   \date{Received xxxx; accepted xxxx}

% \abstract{}{}{}{}{} 
% 5 {} token are mandatory
 
  \abstract
   {Gravitational microlensing is a phenomenon that allows us to observe dark remnants of stellar evolution even if they no longer emit electromagnetic radiation.
   In particular, it can be useful to observe solitary neutron stars or stellar-mass black holes, providing a unique window through which to understand stellar evolution. 
   Obtaining direct mass measurements with this technique requires precise observations of both the change in brightness and the position of the microlensed \changed{star. The European Space Agency's} \gaia satellite can provide both.
   \changed{Using publicly available data from different surveys, we} analysed events published in the \gaia Data Release~3 (\gdr{3}) microlensing catalogue. Here we describe our selection of candidate dark lenses, where we suspect the lens is a white dwarf (WD), a neutron star (NS), a black hole (BH), or a mass-gap object, with a mass in a range between the heaviest NS and the least massive BH.
   We estimated the mass of the lenses using information obtained from the best-fitting microlensing models, the source star, the Galactic model and the expected distribution of the parameters.
   \changed{We found eleven candidates for dark remnants: one WDs, three NS, three mass-gap objects, and four BHs.}
   }

   \maketitle
%
%-------------------------------------------------------------------

\section{Introduction}
There are still many outstanding questions connected with the remnants of stellar evolution.
The most common stellar remnant is a white dwarf, and more than 95 per cent will become one by the end of their life \citep{2001Fontaine}.
Our understanding of white dwarfs was expanded in recent years by \gaia and its superb parallaxes. 
The largest catalogue consists of over 350'000 high-confidence WD candidates, expanding almost ten times the amount of known WDs before \gaia \citep{2021GentileFusilloWD}.
The best catalogue of known pulsars is two orders of magnitude smaller than the one we have for WDs in our Galaxy \citep{2005ManchesterNS}.
Our most limited observational material is on BHs, in particular solitary ones.
Most of the known BHs are linked to binary systems found either through X-ray emission due to accretion of their companions \citep[e.g.][]{2016CorralSantanaBH} or as gravitational wave sources due to their merger \citep[e.g.][]{2019AbbotGWTC-1}. 
Additionally, gravitational wave mergers are detected most frequently in distant galaxies.
Recently, \cite{2022ShenarNoninBH}, \cite{2022ElBadryGBH1}, and \cite{2023ElBadryBH2, 2023Chakrabarti} have reported on BH candidates also detected as non-interacting binary systems. 
However, the only known direct mass measurement for a solitary stellar-mass BH was recently presented for OGLE-2011-BLG-0462/MOA-2011-BLG-191 \citep{2022LamBH, 2022SahuBH, 2022MrozUdalskiGould, 2023LamBHReanalysis} using the gravitational microlensing phenomenon.

Gravitational microlensing is an effect of Einstein’s General Relativity, which occurs when a massive object passes in front of a distant star within the Milky Way or its neighbourhood \citep{1936Einstein, 1986Paczynski}. 
In contrast to strong gravitational lensing, here the separated, deformed images of the source are typically impossible to spatially resolve unless the world’s largest telescopes are used, and only in case of very bright events \citep{2019Dong, 2022Cassan}. 
Instead, what can be observed is a brightening of the source occurring during the event.
Images of the source, though difficult to resolve, are unequally magnified and change position.
This causes a distinctive shift in the centroid of light called astrometric microlensing \citep{2000DominikSahu, 2002BelokurovEvans}.
This effect can be measured with precise enough observatories like \textit{Hubble} Space Telescope (HST) \citep{2017SahuWDAstro}, or \gaia \footnote{\url{https://www.cosmos.esa.int/web/gaia/iow_20210924}}.

Combining both effects allows the mass of the lens $M_\mathrm{L}$ to be measured following \cite{2000Gould}:
\begin{equation}
    M_\mathrm{L} =  \frac{\theta_\mathrm{E}}{\kappa \pi_\mathrm{E}},
\end{equation}
where $\kappa = 4G/c^2\mathrm{au} \approx 8.144 \mathrm{mas}/M_\odot$, $\theta_\mathrm{E}$ is the angular Einstein Radius, which can be measured with astrometric microlensing, and $\pi_\mathrm{E}$ is the microlensing parallax obtained from modelling the time-series photometry. 
A combination of these two effects was used to detect a stellar-mass BH for the first time in \cite{2022LamBH} and \cite{2022SahuBH}, who used astrometric observations from HST and photometric observations from the ground.

However, even without a measurement of the angular Einstein radius, we can still estimate the mass of the lens by employing the Galactic model and expected distributions of lens parameters. 
We can obtain a posterior distribution for the lens mass and distance 
using the microlensing parallax, proper motion measurements, estimated distance to the source, the Galactic model and assumed mass function of stellar remnants \citep[e.g.][]{2016WyrzykowskiBH, 2021MrozWyrzykowski}. 
This method was used for objects observed by the OGLE survey where no Einstein radius information was available, where events exhibited clear parallax signal \citep[e.g.][]{2020WyrzykowskiMandel, 2021MrozBHOGLE}. 
The same technique could also be applied for microlensing events seen by \textit{Gaia} \citep{2016PrustiGaiaMission}, using both archival data and transients detected as part of \textit{Gaia} Science Alerts (GSA) system \citep{2021HodgkinGsa}. 
This paper presents a similar analysis of \textit{Gaia} Data Release~3 (\gdr{3}) microlensing catalogue \citep{2022Wyrzykowski_variUlens}.

This work is split into six sections. 
Section \ref{sec:data} presents the microlensing models compared in this work, the criteria of event pre-selection and the sources of data used for this analysis.
Section \ref{sec:selection} explains the criteria to select events for detailed analysis, while section \ref{sec:analysis} summarises those results.
Section \ref{sec:mass} shows how we estimated the masses and distances to the lenses.
Section \ref{sec:discussion} discusses the obtained results and summarises this work.

\section{Event pre-selection and data}\label{sec:data}
\subsection{Compared models}
In this paper, we focused only on events that could exhibit the microlensing parallax effect, which occurs when the observer changes position during the event.
There are three types of microlensing parallax: annual, terrestrial and space.
The annual microlensing parallax is connected to the Earth's movement around the Sun.
The observer on Earth changes their position during the entire year, which creates distinctive asymmetry and, in some cases, wobbles in the light curve \citep{1992GouldParallaxMACHOs, 1995AlcockFirstParallax, 2023Masokliunas}.
The terrestrial parallax is connected to the different positions of the observatories on Earth.
It is measurable only in the most extreme cases, such as catching a caustic crossing with telescopes on two sites distant from each other \citep{1995HardyWalkerTerrPar, 1996HolzWaldTerrPar, 2009GouldTerPar}.
Finally, space parallax occurs when the event is observed from observatories located on Earth and in space.
When the space observatory is located as far as one au from the Earth, it can cause a significant difference in the amplification and time of the peak of the lens \citep{1966RefsdalSpacePar, 2023SpechtK2C9}. 
It can be also measured if the space observatory is closer but during a caustic crossing \citep{2020Wyrzykowski16aye} or if the event is densely covered.
This is the main mechanism behind the way that the Nancy Grace Roman Space Telescope is going to be used for mass measurements of the observed lenses \citep{2019PennyRoman}.

\gdr{3} microlensing events catalogue contains events which were most likely caused by a single object as an outcome of the used pipeline \citep{2022Wyrzykowski_variUlens}.
% We selected events
All of the events within this catalogue were detected in the Galactic plane, which is a dense field, especially within the Galactic bulge.
This means that we had to include blending when some of the light is coming from the stars near the line of sight towards the source and lens. 
In the case of microlensing, blending also factors in that the lens is luminous in the majority of cases.

We used the following models in our analysis with these parameters:
\begin{itemize}
    \item point source - point lens (PSPL) model without blending, parameterised by $t_0$, $u_0$, $\tE$, $I_0$;
    \item PSPL with blending, parameterised by $t_0$, $u_0$, $\tE$, $I_0$, $f_b$;
    \item PSPL model with parallax effect without blending, parameterised by $t_0$, $u_0$, $\tE$, $I_0$, $\piEN$, $\piEE$;
    \item PSPL model with a parallax effect with blending, parameterised by $t_0$, $u_0$, $\tE$, $\piEN$, $\piEE$, $I_0$, $f_b$;
\end{itemize}
where $t_0$ is the time of the peak of brightness, $u_0$ is the impact parameter at $t_0$, and $\tE$ is the Einstein timescale when the source is crossing the angular Einstein ring.
Microlensing parallax is described by its northern and eastern components $\piEN$ and $\piEE$.
The baseline magnitude of the event is denoted by $I_0$ and the blending parameter is defined as $f_b = \frac{F_b}{F_s + F_b}$, where $F_s$ is the source flux and $F_b$ is the blend flux.

In this work, we used models without blending in the pre-selection stage, and for each event, we fitted models with and without parallax.
We used models with blending when we fitted each event individually.
At this stage, we also fitted models with and without parallax. 
Each event should have at least two best-fitting solutions: PSPL without blending, and PSPL with parallax and without blending.

\subsection{Pre-selection of the candidate events}
The \gaia Data Release~3, or alternatively table \texttt{vari\_microlensing} of the \gdr{3}, contains 363 candidate events.
Many of them do not exhibit second-order effects and are best described by the standard Paczy\'nski model.
We suspected that events with short Einsten timescales are less likely to be affected by the annual movement of the Earth around the Sun.
Thus we selected events with \texttt{paczynski\_0\_te} timescale larger than 50~days. 
This was an arbitrary cut, based on the fact that \gaia produces on average one point per month per source.
An event with an Einstein timescale of 50~days would last more than 100 days, allowing for at least 3 observations during the event.
Additionally, previous studies of candidate parallax events show that in most cases parallax is not detectable for shorter events (see for example \cite{2022Rodriguez} and \cite{2023Zhai}).
After applying this cut, we were left with 204~candidate events to analyse.
% \kkcom{\begin{itemize}
%     \item where data comes from (Gaia DR3 3-band phot, aux data: OGLE, MOA, KMTNet, GSA)
%     \item how did we get initial parameters
%     \item how we selected our sample
% \end{itemize}}
\subsection{Data}
The \cite{2022Wyrzykowski_variUlens} catalogue was built using only \gaia \gmag, \gbp and \grp photometry, but for this work, we utilised data available from other surveys.
In particular, we wanted to include information from microlensing surveys which have better cadence, especially in the Galactic bulge.
We have cross-matched the \gaia sources with the OGLE survey \citep{1992UdalskiOGLE, 2015UdalskiOGLEIV}.
We have found 145 events in common with the public OGLE events.
130 events were published as a part of the OGLE-IV analysis of microlensing optical depth in the Galactic plane \citep{MrozOGLEBulge, MrozOGLEDisc}.
78 events were published were also published as OGLE Early Warning System alerts \citep{2015UdalskiOGLEEws}\footnote{\url{https://ogle.astrouw.edu.pl/ogle4/ews/ews.html}}, overlapping with the 130~events coming from the OGLE-IV papers.
We downloaded all publicly available data. 
If the event was published in OGLE-EWS, \cite{MrozOGLEBulge} or \cite{MrozOGLEDisc}, we used the data shared with the article.
We performed a similar search with MOA survey \citep{1997AbeMOA, 2001BondMOA} using its alert stream and we found 20 events in common.
We found 32 events in common with KMTNet survey public alerts \citep{2014LeeKMTNet, 2016KimKMTNet}\footnote{\url{https://kmtnet.kasi.re.kr/~ulens/}}.
Six events were published by \gaia Science Alerts\footnote{\url{http://gsaweb.ast.cam.ac.uk/alerts/home}}.
These events were published with preliminary photometry and without errors. 
We simulated the errors using the following formula, \citep{2022Wyrzykowski_variUlens}:
\begin{equation} \label{eq:gaia_err}
    \sigma_{G,i} = 
    \begin{cases}
         \sqrt{30}\times10^{0.17\times13.5-5.1}, & ~~ \mathrm{for}~ G_i<13.5~\mathrm{mag},\\
    \sqrt{30}\times10^{0.17\times G_i-5.1}, & ~~ \mathrm{for}~ G_i\geq13.5~\mathrm{mag},\\
    \end{cases}
\end{equation}
where $G_i$ is the $i$-th point in the GSA light curve. 
Since the error bars and photometric data had different properties, they came from different pipelines, and GSA data was created using raw photometric data. 
\gdr{3} light curves were created by the photometric pipeline that was used on all data used for this Data Release and produced the most accurate light curves we have. 
We decided to treat them as a different dataset.
Seven events were found in the publicly available data of the ASAS-SN survey \citep{2014ShappeeASASSN}.
Two were published as alerts: ASASSN-16li and ASASSN-16oe \citep{2016Asassn16oe1, 2016Asassn16oe2}, and one was published as an ATEL \citep{2017AsassnAntiCenter}.
The rest was found in the ASAS-SN Photometric Database \citep{2019JayasingheASASSNVarCat}.
We did not include Zwicky Transient Facility (ZTF) \citep{2019BellmZTF} while cross-matching events, because this survey started after May 2017, which was the end of \gdr{3} timespan.
We did however check for sources appearing in the 9th Data Release of ZTF if a given source brightened only once.
The list of all 204 sources with their names in other surveys is available in Table \ref{tab:xmatch}.

In the case of MOA and KMTNet, we have used the available photometry published in fluxes, instead of magnitudes.
KTMNet is a network of three robotic telescopes, located in Australia, South Africa, and Chile. 
These sites have different weather conditions, and when we used KMTNet DIA photometry, we separated each light curve by the observatory.
For \gaia photometry, we followed \cite{2022Wyrzykowski_variUlens}, and modified the available uncertainties to match the method used to find candidate events.

All data sources listed above were then used either at the preliminary or the detailed event modelling stages or both.
We provide data used for this stage in a machine-readable online archive\footnote{\url{https://github.com/KKruszynska/gdr3_dark_lenses}}.

\section{Selection of candidate events for further analysis}\label{sec:selection}
To find preliminary models, we used the \texttt{MulensModel} package \cite{2019MulensModel} to generate microlensing models, and the \texttt{pyMultiNest} package \citep{2009FerozMultiNest, 2014BuchnerPyMultiNest} to find the best fitting solutions.
To simplify the parameter space explored by the \texttt{pyMultiNest} package, we calculated models without blending.
\texttt{pyMultiNest} provides a Python interface for a nested sampling algorithm which returns the best solutions for probability densities containing multiple modes and degeneracies. 
This made it a perfect tool for comparing models including microlensing parallax.
For the parallax model, we included both the annual and space effects.
\gaia is located in space, and there may be an offset between observatories.

We have recorded the four best solutions for models with and without parallax and compared their $\chi^2$ values. These solutions are available in machine-readable format. 
Using preliminary models we selected events, that:
\begin{itemize}
    \item had Einstein timescale of the best PSPL solution larger than 50~days, and
    \item the difference of $\chi^2$ per degrees-of-freedom of the best PSPL model and the best parallax model should be larger than one ($\chi_{PSPL}^2/\mathrm{dof} - \chi_{Par}^2/\mathrm{dof} > 1$).
\end{itemize}
This way we selected 34~events. 
We removed two events from this sample. 
For the first one (\ugdr{024}), we didn't have a full light curve, and the event did not finish before the end of the \gdr{3} period. 
The second one (\ugdr{178}) turned out to be a binary event, when we inspected the MOA light curve.
We decided to add three additional events, that had ASAS-SN data (\ugdr{023}, \ugdr{032}, and \ugdr{118}).
In these cases, the automatic algorithm struggled to find a correct solution, and we concluded that was caused by the vastly different pixel size of the ASAS-SN, compared to \gaia and other surveys, and the exclusion of blending in fitted models.

% We couldn't obtain sensible preliminary models for events containing ASAS-SN data, so we modelled them separately, using the same pipeline, fitting models with blending.
% This was necessary, because of the vastly different diameter of ASAS-SN telescopes compared to OGLE or Gaia, which in turn manifests by larger pixel size.

% \begin{figure*}
%   \centering
%   \includegraphics[width=\hsize]{Figures/for_paper_ulens-069.png}
%       \caption{Top panel: light curve of the event known as GaiaDR3-ULENS-069, KMT-2016-BLG-1346, and MOA-2015-BLG-158. \gaia \gmag  data is shown in black, \grp in orange, KMTNet $I$-band data is shown in brick red, dark red, and teal, and MOA-red data in light-green. Three best solutions are marked: PSPL without parallax G0 with a black dashed line, and two PSPL with parallax, G+ and G- with red and gray continuous lines respectively. The bracket in top left shows the $\chi^2$ of different solutions.
%       Bottom panel: Residuals of the GSA+ model. Black dashed line marks the GSA0 and GSA+ models difference, while the grey continuous line marks the GSA- and GSA+ models difference.}
%          \label{fig:u069}
%   \end{figure*}

\section{Detailed analysis of selected events}\label{sec:analysis}
We conducted a case-by-case analysis of the 35~events selected in the previous step. 
We used \texttt{MulensModel} to generate the microlensing models and \texttt{emcee} \citep{2013Emcee} to explore the parameter space.
In this step, we used the KMTNet pySIS photometric data in magnitudes.
For KMTNet data, when possible, we pre-processed data, removing any points that had a negative value of FWHM column or with photometric uncertainty larger than 1~magnitude.
For OGLE and MOA events, we used re-processed data coming from the end-of-the-season DIA photometric reduction pipelines \citep{2015UdalskiOGLEIV, 2001BondMOA}.
We also applied a correction procedure for uncertainties following \cite{2016Skowron} to OGLE data from bulge fields.
Finally, we re-scaled the photometric uncertainties using this formula:
\begin{equation}
    \sigma'_{n,i} = k_n \sigma_{n,i},
\end{equation}
where $\sigma'_{n,i}$ is the re-scaled $i$-th uncertainty of the $n$-th telescope's light-curve, $k_n$ is the scale factor for the $n$-th telescope, and $\sigma_{n,i}$ is the original $i$-th uncertainty of the $n$-th telescope's light-curve.
We obtained the scale factors in the following manner:
\begin{enumerate}
    \item we fitted preliminary PSPL models with and without parallax;
    \item we selected a model with the smallest $\chi^2$ value;
    \item using this model we used it as a starting point, we ran an MCMC fit with scale factors as one of the fit parameters.
\end{enumerate}
If the scale factor of the median solution found in the final step exceeded 1.0 for a given telescope, we used this value and then we re-fitted the PSPL models with and without parallax. 
For events \ugdr{003}, \ugdr{032}, \ugdr{118}, \ugdr{196}, and \ugdr{284} we had to use the PSPL without parallax model instead of the best model to find the scale factors.
For event \ugdr{025}, \ugdr{143}, we couldn't find scale factors due to poor event coverage.
We report the values for each scale factor in \ref{tab:fit_data}.
For some events, we had to perform an outlier removal procedure.
We have done this before applying the uncertainty re-scaling.
We used the best-fitting preliminary model (step 2 of the uncertainty re-scaling procedure).
Then we removed all data points outside the 3 $\sigma$ range of the residuals from the preliminary model for a given light curve. 
We marked those light curves in bold in Table \ref{tab:fit_data}.
In some cases, we had to remove certain light curves, because they were too noisy, or carried little information about the event.
We marked those data sets with a strike-through text in Table \ref{tab:fit_data}.
Table \ref{tab:fit_data} provides the name of the fitted event and a list of data sets with the amount of data points for each light curve.
Table \ref{tab:fit_best_model} presents the median values of the posterior distributions (PDFs) obtained for best-fitting solutions.
KMTNet data for event \ugdr{047}/KMT-2015-BLG-0157 revealed features around the peak, for which a PSPL model was insufficient in characterising, so it was excluded from further analysis.
\ugdr{284} had a large parallax value, which means other effects should be included. We excluded that event from further analysis.
\ugdr{057} had only three points at magnification, so we could only find a non-parallax solution.
We named the solutions with the following convention: "{\gaia}DR3-AAA-BC", where AAA is the number assigned to the event in \cite{2022Wyrzykowski_variUlens}, B is a string of letters denoting which type of \gaia data was used: "G" for events where we used the \gdr{3} photometry, and "GSA" for events where we opted for a \gaia Science Alerts lightcurve, finally C is a sign of the $u_\mathrm{0}$ of the solution ("+" for positive and "-" for negative). If there was more than one solution with the same $u_\mathrm{0}$ sign, we numbered them starting from one.

To select the dark lens candidate sample, we applied the following criteria to the modelled events:
\begin{itemize}
    \item the blending parameter in \gmag band was smaller than 0.3,
    \item the $\piE$ was not consistent with zero in the three-sigma range,
    \item the $\chi^2$ of the parallax solution was smaller than the $\chi^2$ of the non-parallax solution.
\end{itemize}
This way we obtained \changed{14}~events, where at least one solution passed those criteria. 
For these events, we then estimated the lens distance and mass.
From the remaining events, a non-parallax model better described two, seven didn't pass the blending parameter criterion, and six did not pass the $\piE$ criterion. 
We found four events that passed the blending and $\chi^2$ criterion but failed the $\piE$ criterion in the calculated three sigma, however, their $\piE$ distribution was inconsistent with 0.
We analysed these solutions further but displayed their results in a different table.

\section{Source stars}\label{sec:source_star}

To determine the properties of the lens, we have to determine the properties of the source. 
Ideally, we would obtain the distance to the source, but that isn't always possible. 
Instead, we decided to find the angular stellar radius of the source $\theta_*$ and use it as a prior during lens mass and distance estimation.
We followed different procedures, depending on the event location and available information.

We could use the colour-magnitude diagrams (CMDs) calibrated to the OGLE-III data for events with MOA data (\ugdr{035}, \ugdr{069}, \ugdr{073}, \ugdr{088}, \ugdr{155}, \ugdr{343}, \ugdr{353}). 
First, we determined the red clump centre (RCG) location following the procedure outlined in \cite{2013Nataf}.
We use the de-reddened RCG distance modulus determined in \cite{2013Nataf} for each event\footnote{\url{https://ogle.astrouw.edu.pl/cgi-ogle/getext.py}}, and find the reddened distance modulus of the RCG using fitted position on the CMD and absolute magnitude of the RCG $M_{I,\mathrm{RCG}} = (-0.12, 1.06)~\mathrm{mag}$ \cite{2013Nataf, 2013Bensby}.
Then we calculate the extinction in $A_I$ and $A_V$ and use it to find the de-reddened magnitude and colour of the source. 
In the case of \ugdr{353}, the blending parameter was negative, so instead of using calculated source magnitude, we used the baseline magnitudes to determine source brightness and colour.
Finally, we use these values to determine the angular stellar radius of the source star using relations from \cite{2018Adams}.

Other sources were more difficult.
If the source was located towards the Galatic Bulge, we assumed that the Bulge is located 8.1~kpc and extends for 2.4~kpc.
We used this value to determine the de-reddened distance modulus to the red clump centre. 
We constructed a CMD in $V$ and $I$ data using \gdr{3} sources. 
We selected sources within 30' of the event, that had the Renormalized Unit Weight Error (RUWE) parameter smaller than 1.4, astrometric parallax error not larger than 20\% of the measured value, and that had available GSP-Phot solutions \citep{2023Andrae}. Then, we transformed their colours into $V$ and $I$ bands following relations from \cite{2022GaiaDocPhot} Section 5.5.1\footnote{\url{https://gea.esac.esa.int/archive/documentation/GDR3/Data_processing/chap_cu5pho/cu5pho_sec_photSystem/cu5pho_ssec_photRelations.html}}.
We determined the position of the RGC following the procedure in \cite{2013Nataf}, and used the calculated distance modulus to find the extinction in $I$ and $V$ bands.
Finally, we found the $\theta_*$ using relation from \cite{2018Adams}.
We used this method for events \ugdr{025}, \ugdr{089}, \ugdr{142}, and \ugdr{270}.
For many of these events, some information was missing.
When baseline magnitude and blending in \gbp and/or \grp filter, we used blending parameter in \gmag (for \gbp) or in $I$ (for \grp) bands and \gdr{3} entry for this source to determine the missing brightness for the $\theta_*$ estimation. We couldn't use this method for events  \ugdr{331} and \ugdr{363}, because they were too dim compared to the data with GSP-Phot entries to infer the extinction.

Finally, for events located towards the Galactic disc, we couldn't determine the de-reddened distance modulus towards the RCG, and therefore $\theta_*$.
This affected events \ugdr{103}, \ugdr{118}, and \ugdr{259}.

All determined values can be found in Table~\ref{tab:source_prop}.

\section{Estimating lens parameters of the candidate dark events}\label{sec:mass}
We used the same approach as in \cite{2016WyrzykowskiBH}, \cite{2021MrozWyrzykowski} and \cite{2022KruszynskaGaia18cbf} to estimate the mass and distance to the lens. 
We dubbed it the \texttt{DarkLensCode}\footnote{\url{https://github.com/BHTOM-Team/DarkLensCode}}.
and we explained this method in greater detail in \cite{2024Howil}.
The \texttt{DarkLensCode} was used to find the posterior distribution of lens distance and lens mass, using PDFs of the photometric model parameters, and the Galactic model. The final estimates are the median values of obtained mass and distance PDFs.
Here we will focus on the presentation and the resulting mass and distance estimates.
We present the results in Tables~\ref{tab:lens_mass} and, Table~\ref{tab:lens_mass_edge}.

If we found more than one solution passed the criteria outlined in Section \ref{sec:analysis}, we analysed them separately providing mass estimates for each solution.
% , we combined the resulting posterior distribution into one, randomly selecting 50~000~points for each solution.

The extinction $A_G$ was calculated following a method similar to one outlined in \cite{2019Fukui}, but we used \gdr{3} data instead.
We selected all sources within a 30' radius with a Renormalised Unit Weight Error parameter smaller than 1.4, parallax uncertainty smaller than 20\% of the measured value, and available GSP-Phot solutions \citep{2023Andrae}.
Then we calculated the mean and standard deviation of the \texttt{ag\_gspphot} in 50 pc bins and fitted a fourth-order polynomial.
We used the fitted polynomial as a function of extinction depending on the distance towards the lens or the source.
If the distance to the lens or source was larger than 8~kpc, we used the calculated extinction value $A_G$ at 8kpc.

We did not know the distance to the source, so we assumed different maximum and minimum ranges.
For events located towards the bulge, we initially assumed that the distance can be between 1~kpc and 12~kpc. 
For events located towards the disc, we chose the distance between 0.1~kpc and 8~kpc (\ugdr{118}) or 10~kpc (\ugdr{259}).
When available, we used $\theta_*$ derived in Section~\ref{sec:source_star}. 
First, we randomly selected a distance to the source from the range described above and calculated the source radius $R_\mathrm{S}$. 
In the next step, we found the absolute magnitude of the source star.
The procedure depended on the position of the source star in the CMD.

If the source star position on the CMD was located in the main sequence, we assumed the star was a dwarf. 
Using the tables from \cite{2013PecautMamajek} provided on the author's website\footnote{\url{https://www.pas.rochester.edu/~emamajek/}} we found the corresponding value of absolute magnitude in \gmag band.

If the source star position on the CMD was located in the red giant clump, we assumed the star was a giant.
For the absolute magnitude determination, we followed the information contained in \cite{2021vanBelle}.
First, we found the $(V_\mathrm{0} - K_\mathrm{0})$ value corresponding to the radius based on the inverted relation presented in the paper with coefficients coming from Table~16.
Then we used Equation~4 from \cite{2021vanBelle} to find the effective temperature $T_\mathrm{eff}$ of the source star.
We used the well-known relation $L_\mathrm{bol}=4\pi R^2_\mathrm{S}\sigma_\mathrm{SB}T^4_\mathrm{eff}$ to find the bolometric luminosity of the source star.
To find the bolometric correction $BC_\mathrm{G}$ in \gmag band, we followed the recipe provided in \cite{2018DR2Manitega_doc}, Chapter~8.3.3.
Finally, we could derive the absolute magnitude in \gmag following Equation~8.6 from \cite{2018DR2Manitega_doc}.

We then found the extinction $A_G$ and the observed magnitude of the source star at a selected distance. 
If the absolute value of the difference between the calculated source magnitude and the observed source magnitude from the microlensing model was smaller than the sum of the source magnitude uncertainty from the model and the derived source magnitude uncertainty, we accepted that source distance value.

We used three mass functions as lens mass priors: \cite{2001Kroupa} function that describes stars:
\begin{equation}
    f(M) \sim 
    \begin{cases}
        M^{-0.3}, & ~~ M \leq 0.08 \Msun,\\
        M^{-1.3}, & ~~  0.08 \Msun < M \leq 0.5,\\
        M^{-2.3}, & ~~  0.5 \Msun < M < 150 \Msun,\\
    \end{cases}
\end{equation}
\cite{2021MrozBHOGLE} function that describes solitary dark remnants in our Galaxy:
\begin{equation}
    f(M) \sim 
    \begin{cases}
        M^{0.51}, & ~~ M \leq 1.0 \Msun,\\
        M^{-0.83}, & ~~  1.0 \Msun < M < 100 \Msun,\\
    \end{cases}
\end{equation}
and a $f(M) \sim M^{-1}$ corresponding to applying no prior on the lens function.
The reported values of the lens mass and distance are median values of the posterior distribution, while their uncertainty is represented by the 16-th and 84-th quantiles.

We noticed that the mass function greatly affects the lens mass estimate. 
Using the \cite{2001Kroupa} mass function results in lighter lenses at greater distances, and in turn more likely MS stars.
In contrast, the \cite{2021MrozBHOGLE} mass function produced more massive lenses at closer distances. 
This is because the \cite{2001Kroupa} mass function is steeper and \changed{less likely to produce} massive lenses.

We compared the brightness of the blend from the microlensing model to the brightness of an MS star at an estimated distance from the lens. 
We summed the number of solutions where the brightness of the blend was smaller than the MS brightness and divided this number by the number of all solutions. 
The resulting number was interpreted as the probability that the lens is dark and not an MS star.

%The extinction $A_G$ used for the calculation of the observed magnitude of an MS was taken, when possible, from \gdr{3} (\texttt{ag\_phot} value). 
%Otherwise, we calculated $A_G$ using the reddening maps of \cite{2011SchlaflyRedd} and a relation of $E(B-V)$ with $A_G$ from \cite{2019WangChenExt}.
%We compared the $A_G$ value to extinction in $u'$, $g'$, $r'$, and $i'$ filters from \cite{2011SchlaflyRedd} to confirm its validity.
%When we included extinction, it resulted in a lower probability for a dark lens. 
All input parameters are available in machine-readable form in an online archive attached to this paper.

\section{Discussion and conclusions}\label{sec:discussion}

We found in total \changed{11}~lenses for which the probability for the dark lens scenario for at least one solution exceeded 80\% when we looked only at the \cite{2001Kroupa} mass function. \changed{Eight} of them passed all of the criteria imported in Section~\ref{sec:analysis} (\ugdr{025}, \ugdr{035}, \changed{\ugdr{069}}, \ugdr{073}, \changed{\ugdr{088}, \ugdr{155},} \ugdr{343}, and \ugdr{353}).
Among the solutions that didn't pass the $\piE$ criterion, we found three more candidates (\ugdr{103}, \ugdr{212}, \ugdr{331}, \changed{and \ugdr{155}, but we accounted for this event in the first group}).
% We can expand this list by six additional events when demanding that probability only the probability $P_0$ is larger than 80\% (\ugdr{
% 155}, \ugdr{212}, \ugdr{259}, \ugdr{270}, \ugdr{363}).
% The probability $P_0$ was calculated without including the extinction.
All but one of these events have Galactic coordinates towards the Galactic Centre.
The estimated distance for \changed{16}~solutions (\changed{eight} events) would suggest that they are located in the Galactic disc, rather than the Galactic bulge. 
For \changed{three} solutions (\changed{two} events), the estimated lens distance suggests a Galactic bulge lens.
\changed{Three events seem to have the lens located within one kiloparsec from Earth.}
One event located towards the Galactic disc seems to have a lens no closer than 3.1~kpc, which means it would belong to the Scutum-Centaurus Arm of the Milky Way.
%This would mean, that the extinction should be smaller than the value used in this work.
%This, in turn, means, that it's less likely that the object would be an MS star, since the true brightness of an MS star $G_\mathrm{MS, true}$ should lie between two marginal values calculated in this paper, namely $G_\mathrm{MS, 0} < G_\mathrm{MS, true} <  G_\mathrm{MS, A}$.
%Additionally, these eight lenses would belong to the Galactic disc population and five of them are located in the nearest 2.5~kpc radius.
%Only one lens, \ugdr{212}, seems to be located near the Galactic bulge and would belong to that population.

If we assumed that all of those \changed{11}~events are dark lenses, and instead follow a \cite{2021MrozBHOGLE} mass function, we would end up with the majority of events with masses in the range of \changed{mass-gap objects and black holes}. 
\changed{One event, \ugdr{035}, has mass consistent with a white dwarf, and three (\ugdr{025}, \ugdr{155}, \ugdr{212}) are consistent with a neutron star.}
\changed{Three events, \ugdr{069}, \ugdr{103} and \ugdr{343} overlap with the first mass gap in the one-sigma range.
Four objects have larger masses (\ugdr{073}, \ugdr{088},  \ugdr{331}, \ugdr{353}) are consistent with BHs.
All BH candidates have large uncertainties. 
There are issues with mass estimates for two of them: for \ugdr{073} we struggled with finding the correct position of the RGC, which could result in the wrong source distance, leading to the wrong lens distance.
\ugdr{331} belongs to the group which didn't pass the $\piE$ criterion in Section~\ref{sec:analysis}.}

\begin{figure*}
  \centering
  \includegraphics[width=17cm]{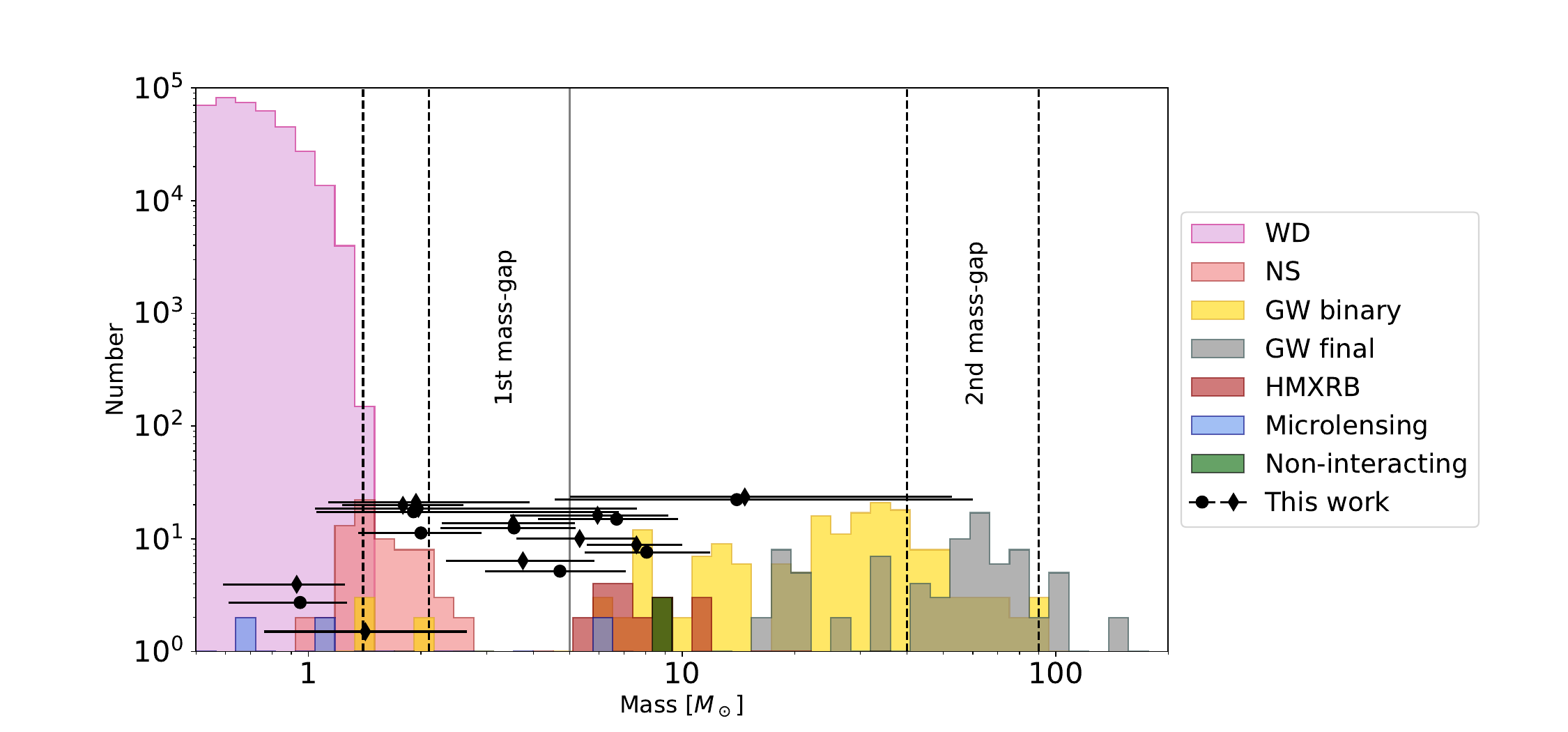}
      \caption{%\changed{This plot was updated.} 
      Distribution of known masses of WDs, NS and BHs.
      Light pink marks WDs known from \gaia \cite{2021GentileFusilloWD}.
      In light red we marked NS with known masses coming from John Antoniadis's catalogue \citep{2012Lattimer, 2013AntonidasNS}.
      Objects found by gravitational wave detectors were marked in yellow \citep{2019AbbotGWTC-1, 2021AbbottGWTC-2, 2021AbbottGWTC-2.1, 2021AbbottGWTC-3, 2021AbbottOpenData}. 
      In red we marked high mass x-ray binaries \citep{2007Orosz, 2007ValBaker, 2009Orosz, 2014Orosz, 2016CorralSantanaBH, 2021MillerJonesCygX1}. 
      In light blue, we marked candidates for dark remnants found by microlensing \citep{2017SahuWDAstro, 2022KaczmarekVVV, 2022KruszynskaGaia18cbf, 2022Jablonska, 2023McGill}, including \cite{2023LamBHReanalysis}.
      In olive, we marked non-interacting dark remnants \citep{2022ShenarNoninBH, 2022ElBadryGBH1, 2023ElBadryBH2, 2023Chakrabarti, 2024ElBadry, 2024A&APanuzzo}. 
      Black dots mark masses of objects known from this work.
      Each solution for the event is shown separately. We marked solutions with positive $u_\mathrm{0}$ with circles, and negative with diamonds. 
      Dashed, vertical lines mark different mass thresholds: the leftmost is the Chandrasekhar mass limit, second to the left is the Tolman-Volkoff-Oppenheimer limit, and two rightmost mark the limits for the theoretical pair-instability supernovae region \citep{2020Farmer}.
      A solid vertical line marks the conventional limit of $5\,\Msun$ for the lightest BHs. 
      Data used to create this plot can be found in \url{https://github.com/KKruszynska/dark_lens_plots/tree/main}.
              }
         \label{fig:masses}
  \end{figure*}

We present these estimates, comparing them to dark remnant mass estimates found through other methods in Figure \ref{fig:masses}.

\begin{figure*}
  \centering
  \includegraphics[width=17cm]{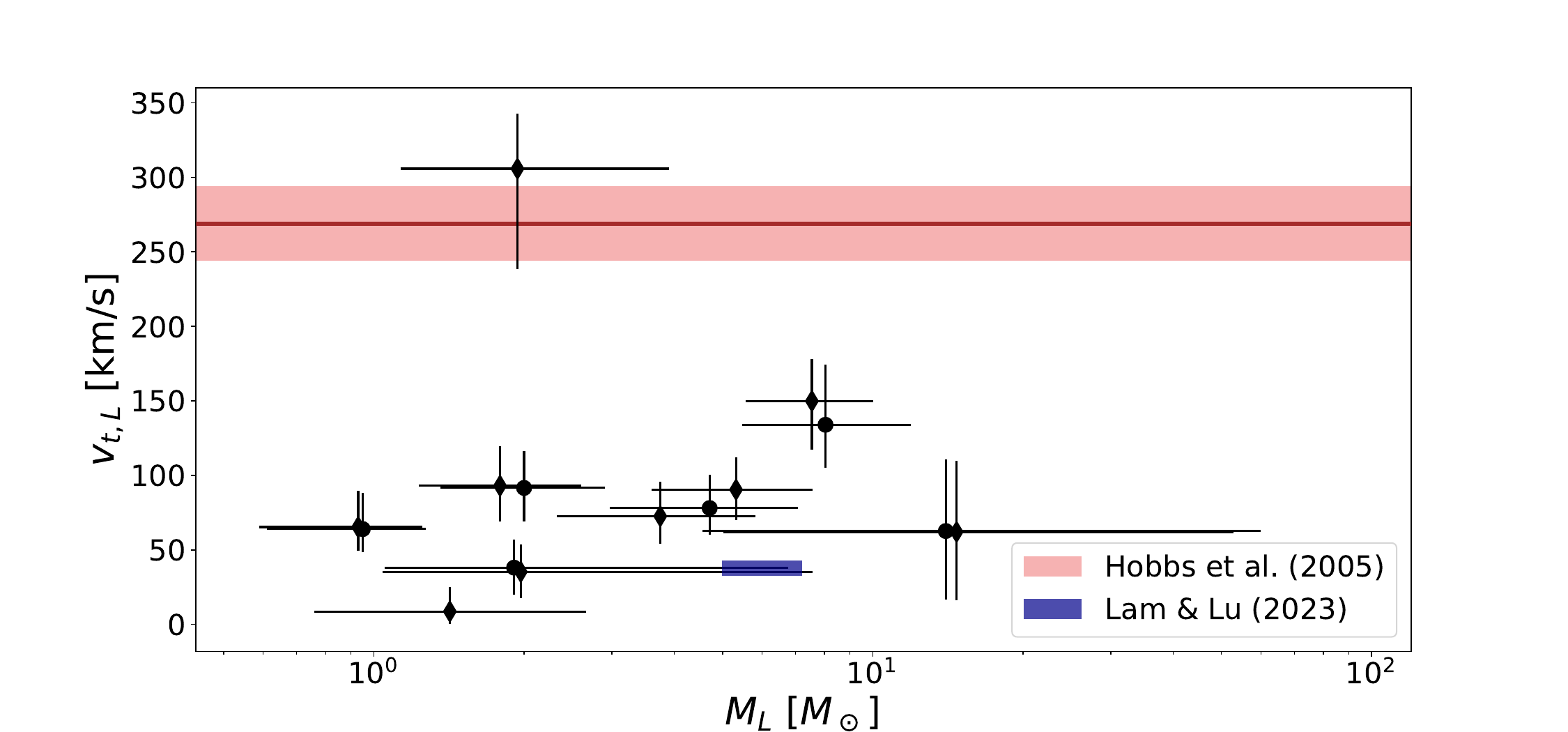}
      \caption{%\changed{This plot was updated.} 
    Transverse velocities $v_\mathrm{t, L}$ estimated for \changed{nine} candidate events using \texttt{DarkLensCode} and \gaia proper motion measurements. $M_\mathrm{L}$ is the mass of the lens. Black dots represent estimates of the masses and velocities of the \changed{nine} candidates with error bars as lines, and each solution is plotted separately. We marked solutions with positive $u_\mathrm{0}$ with circles, and negative with diamonds. The red line represents the median transverse velocity of NS from \cite{2005HobbsNS}, with a light red rectangle representing their dispersion. The dark blue rectangle represents the mass and transverse velocity of the BH from \cite{2023LamBHReanalysis}.
              }
         \label{fig:vt}
  \end{figure*}

In the case of events where \gaia reported a proper motion, we calculated the transverse velocity.
We present results for selected \changed{nine} events in Figure \ref{fig:vt}.
\changed{There is one high-velocity events: \ugdr{212}.
This event has a relatively short Einstein time scale of  56~days.
The estimated mass is consistent with an NS. 
The proper motion found in \gdr{3} is not out of the ordinary, so this means that the lens would have to move fast to justify the relative proper motion and the resulting Einstein time scale.}

All \changed{nine} objects seem to separate into two groups: high-velocity, similar to the NS velocity from \cite{2005HobbsNS}, and low-velocity, similar to the velocity of a solitary BH from \cite{2023LamBHReanalysis}.
The velocity doesn't seem to depend on the lens mass, but the error bars are large and it is hard to draw conclusive statements.
Moreover, here we are most likely getting an estimate of a velocity coming from a Galactic prior.
\changed{More accurate estimates will be possible once we obtain an astrometric time series for these events.}

Here we also measure only the most likely mass of candidate events.
We cannot, however, confirm their nature as WDs, NS or BHs until we perform additional observations in other ranges of electromagnetic radiation, especially X-rays and UV.
They could be some unusual objects, such as quark stars, or
products of primordial black hole (PBH) mergers with other objects.
\cite{2018ABW} and \cite{2022AbramowiczPBH} show a mechanism, that could produce a low-mass BH from a moon-mass PBH ($10^{25}\,\mathrm{g} > M_\mathrm{PBH} > 10^{17}\,\mathrm{g}$) collision with an NS.
Such low-mass BH would be in the mass range of an NS.
% This mass range was excluded by the microlensing observations of the Magellanic Clouds \kkcom{cite proper macho papers}.

It is worth noting, that these are only candidates for dark lenses, and their mass measurement will be possible only when we will include astrometric time series.
This data will be available only with \gdr{4}, no sooner than the end of 2025.
This is an exciting prospect, as \gaia may allow us to observe previously unseen stellar populations.
This becomes even more promising with the approaching start of the Vera C. Rubin Observatory and its Legacy Survey of Space and Time \citep{2019IvezicRubin}, as well as the launch of the Roman mission \citep{ 2015SpergelRoman, 2019AkesonRoman}.
Rubin will allow us to select long-duration microlensing events from the entire Galactic plane, while Roman can provide us with high-cadence astrometric and photometric observations in the Galactic bulge.

\begin{acknowledgements}
KK would like to thank Monika Sitek, Drs. Etienne Bachelet, Mariusz Gromadzki, Przemek Mr{\'o}z, Radek Poleski, Milena Ratajczak, Rachel Street, Pawe{\l} Zieli{\'n}ski, Przemys{\l}aw Miko{\l}ajczyk, as well as Profs. Micha{\l} Bejger, Wojciech Hellwing, Katarzyna Ma{\l}ek, and {\L}ukasz Stawarz.
% Harmonia and Daina    
This work was supported from the Polish NCN grants: Harmonia No. 2018/30/M/ST9/00311, Daina No. 2017/27/L/ST9/03221, and 
%Kartezjusz!!! 
NCBiR grant within POWER program nr POWR.03.02.00-00-l001/16-00.
% ORP acknowledgment
LW acknowledges MNiSW grant DIR/WK/2018/12 and funding from the European Union's Horizon 2020 research and innovation program under grant agreement No. 101004719 (OPTICON-RadioNET Pilot, ORP).
The MOA project is supported by JSPS KAKENHI Grant Number JP24253004, JP26247023, JP16H06287 \changed{and JP22H00153.}
%YT acknowledges the support of DFG priority program SPP 1992 “Exploring the Diversity of Extrasolar Planets” (TS 356/3-1).
This work has made use of data from the European Space Agency (ESA) mission {\it Gaia} (\url{https://www.cosmos.esa.int/gaia}), processed by the {\it Gaia} Data Processing and Analysis Consortium (DPAC, \url{https://www.cosmos.esa.int/web/gaia/dpac/consortium}). Funding for the DPAC has been provided by national institutions, in particular, the institutions participating in the {\it Gaia} Multilateral Agreement.
We acknowledge ESA \gaia, DPAC and the Photometric Science Alerts Team (\url{http://gsaweb.ast.cam.ac.uk/alerts}).
This research has made use of publicly available data 
(https://kmtnet.kasi.re.kr/ulens/) from the KMTNet system
operated by the Korea Astronomy and Space Science Institute
(KASI) at three host sites of CTIO in Chile, SAAO in South
Africa, and SSO in Australia. Data transfer from the host site to KASI was supported by the Korea Research Environment Open NETwork (KREONET).
\end{acknowledgements}

\bibliographystyle{aa}
\bibliography{biblio}

\begin{figure*}
  \centering
  % \sidecaption
  \includegraphics[width=12cm]{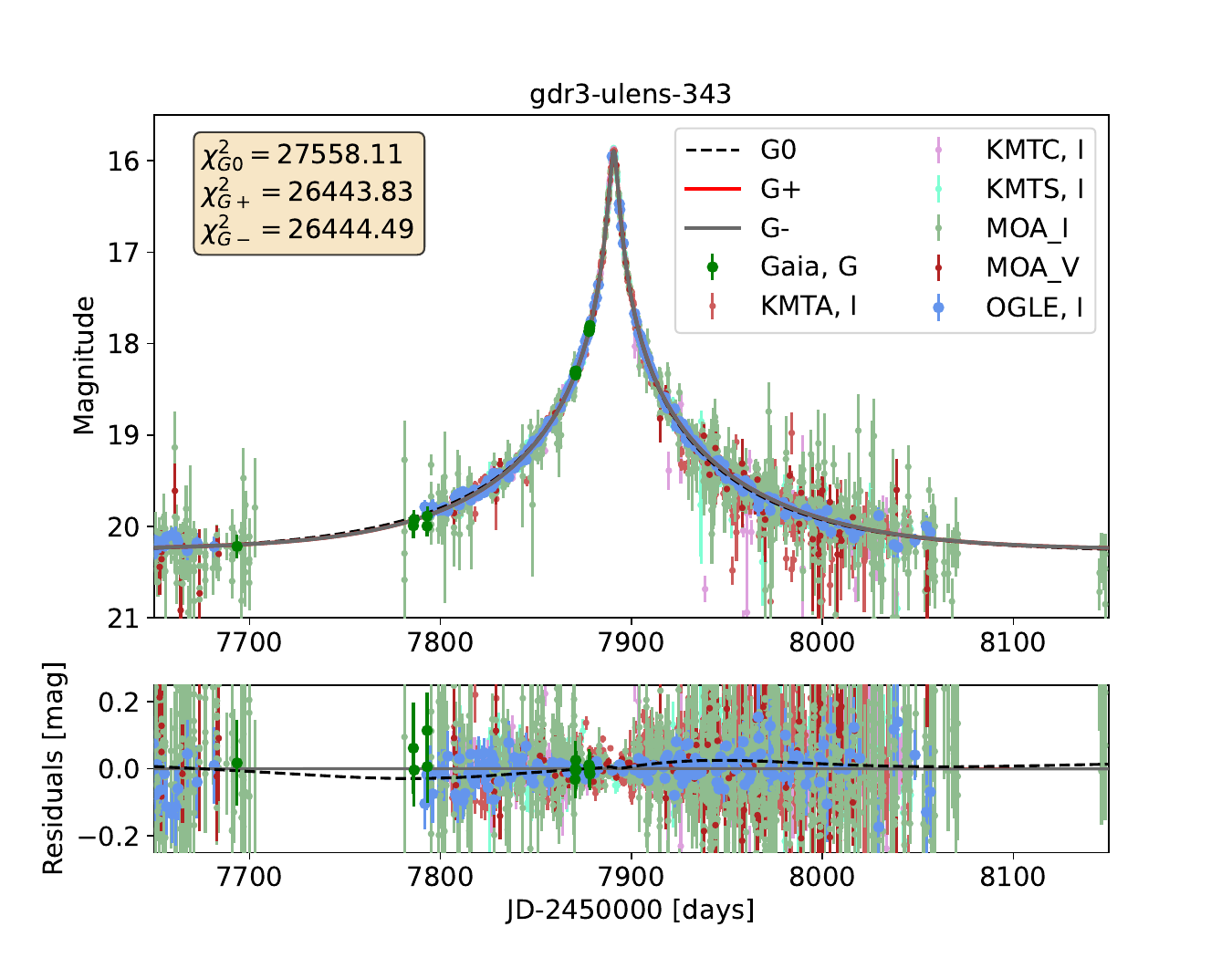}
      \caption{%\changed{This plot was changed, and a different event was chosen. The caption was updated.} 
      Top panel: light curve of the event known as GaiaDR3-ULENS-343, BLG502.29.100629 \cite{MrozOGLEBulge}, OGLE-2017-BLG-0095, MOA-2017-BLG-160, KMT-2017-BLG-1123. \gaia \gmag data is shown in green, \grp in red, KMTNet data is shown in dark red, violet and aqua for the South African Astronomical Observatory (KMTA), Cerro Tololo Inter-American Observatory (KMTC), and Siding Springs Observatory (KMTS), MOA I and V band data are shown in light green and dark red respectively, and OGLE in light-blue. The four solutions are marked: PSPL without parallax G0 with a black dashed line, and two PSPLs with parallax, G+ and G- with red and grey continuous lines respectively. The bracket in the top left shows the $\chi^2$ of different solutions.
      Bottom panel: Residuals of the G+ model. A Black dashed line marks the G0 and G+ models difference, while the dark grey continuous line marks the G+ and G- models difference respectively.}
     \label{fig:u343}
  \end{figure*}

\begin{figure*}
  \centering
  % \sidecaption
  \includegraphics[width=12cm]{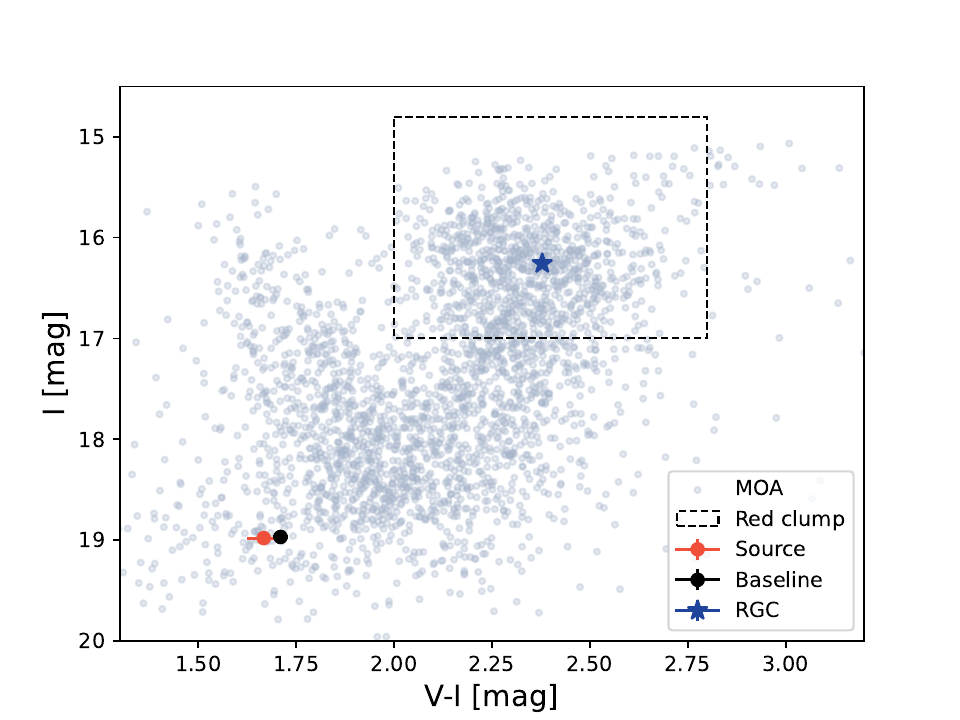}
      \caption{%\changed{This is a new plot.} 
      Colour-magnitude diagram based using MOA data calibrated to OGLE-III catalogue for event \ugdr{343}. MOA data is displayed in light blue dots. The red dot marks the source position, black dot marks the blended baseline source colour and magnitude. Dashed lines mark the region that we used for estimating RGC, and the dark blue star represents the found RGC position.}
         \label{fig:cmd155+}
  \end{figure*}

\begin{appendix}
\section{Additional tables}
In this appendix, we provide five tables:
\begin{itemize}
\item Table \ref{tab:xmatch}, which is the result of the cross-match between the preliminary sample of 204 \gdr{3} microlensing events and other surveys;
\item Table \ref{tab:fit_data}, which is the list of data sets used to obtain the final models for each of the 35~analysed events;
\item Table \ref{tab:source_prop}, which contains information about the source star;
\item  Table \ref{tab:lens_mass} with the lens mass and distance estimates of the 14~candidate dark lens microlensing events;
\item Table \ref{tab:lens_mass_edge} with the lens mass and distance estimates of the five candidate dark lens microlensing events, that didn't pass the $\piE$ criterion, but were chosen to be analysed further.
\item Table \ref{tab:fit_best_model}, which is the list of parameters of best-fitting solutions of the 35~analysed events. This paper only provides the baseline magnitude and blending parameter for the \gmag-band. A full, machine-readable version of this table, with all parameters, is available online.
\end{itemize}

\begin{table}
\begin{center}
\caption{\label{tab:xmatch} Table of the results of cross-match between 204 analysed \gdr{3} events with other surveys. 
A full, machine-readable\\version of this table is available at the CDS.
}
\begin{tabular}{c c c c}
\hline
\addlinespace
{\gaia}DR3- & RA & Dec & \multirow{2}{*}{Remarks}\\
ULENS- & [deg] & [deg] & \\
\addlinespace
\hline
\hline
\addlinespace
002 & 258.0717 & -16.8677 & AP29744086 \\
\addlinespace
003 & 284.4367 & -20.7758 & AP28506888 \\
\addlinespace
\multirow{2}{*}{007} & \multirow{2}{*}{270.2540} & \multirow{2}{*}{-32.6896} & OGLE-2015-BLG-0064 \\
& & & AP27808437 \\
\addlinespace
013 & 274.7611 & -27.7294 & OGLE-2015-BLG-1755 \\
\addlinespace
... & ... & ... & ...\\
\addlinespace
\multirow{3}{*}{363} & \multirow{3}{*}{266.9194} & \multirow{3}{*}{-25.0204} & BLG633.01.52040 \\
& & & OGLE-2017-BLG-0116 \\
& & & KMT-2017-BLG-1029 \\
\addlinespace
\hline
\end{tabular}
\end{center}
\end{table}

\onecolumn
\begin{ThreePartTable}
\begin{TableNotes}
\item{\textbf{Notes:} The table contains the event name, the name of the survey and the filter, the number of points used to obtain the best models, and the scale factor $k$ used to re-scale the photometric uncertainties. Light curves not used in the modelling process are marked with a strike-through. Light curves that were cleaned using the procedure outlined in Section~\ref{sec:analysis} are marked in bold.}
\end{TableNotes}
\LTcapwidth=\textwidth
\begin{longtable}{c c c c c c c c}
\caption{\label{tab:fit_data} %\changed{Values in this table were changed. New columns were added.} 
The list of data sets used to obtain the final models for each of the 35 analysed events. The description of the columns is at the end of the table.}\\
\hline
\addlinespace
{\gaia}DR3- & Survey, & Num. & \multirow{2}{*}{$k$} & {\gaia}DR3- & Survey, & Num. & \multirow{2}{*}{$k$}\\
ULENS- & filter & of points & & ULENS- & filter &  of points & \\
\addlinespace
\hline
\hline
\addlinespace
\endfirsthead
\caption{continued.}\\
\hline
\addlinespace
{\gaia}DR3- & Survey, & Num. & \multirow{2}{*}{$k$} & {\gaia}DR3- & Survey, & Num. & \multirow{2}{*}{$k$}\\
ULENS- & filter & of points & & ULENS- & filter &  of points & \\
\addlinespace
\hline
\hline
\addlinespace
\endhead
\addlinespace
\hline
\endfoot
\addlinespace
\hline
\insertTableNotes
\endlastfoot
\multirow{7}{*}{003} & & & & \multirow{7}{*}{018} & \gaia, \gmag & 24 & 1.00 \\
& & & & & \gaia, \gbp & 22 & 1.00 \\
& \gaia, \gmag & 27 & 1.00 & & \gaia, \grp & 22 & 3.13 \\
& \gaia, \gbp & 27 & 1.00 & & OGLE, I & 2780 & 1.02 \\
& \gaia, \grp & 27 & 1.00 & & \textbf{ KMTNet SAAO, I} & \textbf{563} & \textbf{4.47} \\
& ASASSN, V & 84 & 1.79 & &\textbf{ KMTNet CTIO, I} & \textbf{840} & \textbf{ 1.97} \\
& & & & & \textbf{KMTNet SSO, I} & \textbf{750} & \textbf{2.11} \\
\addlinespace \hline  \addlinespace
\multirow{5}{*}{023} & \gaia, \gmag & 64 & 1.00 & \multirow{5}{*}{025} & \gaia, \gmag & 33 & -- \\
& \textbf{\gaia, \gbp} & \textbf{62} & \textbf{1.00} & & \gaia, \gbp & 32 & -- \\
& \textbf{\gaia, \grp} & \textbf{60} & \textbf{1.00} & & \gaia, \grp & 32 & -- \\
& OGLE, I & 127 & 1.50 & & OGLE, I & 99 & -- \\
& \textbf{ASASSN, V} & \textbf{435} & \textbf{1.00} & & & \\
\addlinespace \hline  \addlinespace
\multirow{6}{*}{032} & & & & \multirow{6}{*}{035} & \gaia, \gmag & 36 & 1.00 \\
& \gaia, \gmag & 40 & 1.00 & & \gaia, \gbp & 33 & 1.00 \\
& \gaia, \gbp & 38 & 1.00 & & \gaia, \grp & 33 & 1.00 \\
& \gaia, \grp & 37 & 1.00 & & OGLE, I & 1063 & 1.00 \\
& ASASSN, V & 308 & 2.09 & &  MOA, I & 19531 & 1.63 \\
& & & & & MOA, V & 612 & 1.89 \\
\addlinespace \hline  \addlinespace
\multirow{6}{*}{047} & \gaia, \gmag & 37 & 1.00 
 & \multirow{6}{*}{57} &  &  & \\
& \textbf{\gaia, \gbp} & \textbf{34} & \textbf{1.00} & & \gaia, \gmag & 24 & -- \\
& \gaia, \grp & 35 & 1.00 & & \gaia, \gbp & 24 & -- \\
& \textbf{KMTNet SAAO, I} & \textbf{2051} & \textbf{6.69} & & \gaia, \gbp & 23 & -- \\
& \textbf{KMTNet CTIO, I} & \textbf{3871} & \textbf{1.00} & &  &  & \\
& \textbf{KMTNet SSO, I} & \textbf{2726} & \textbf{4.77} & & & & \\
\addlinespace \hline  \addlinespace
\multirow{8}{*}{069} & \gaia, \gmag & 36 & 1.00 & \multirow{8}{*}{073} & \gaia, \gmag & 22 & 1.00 \\
& \gaia, \gbp & 31 & 1.00 & & \sout{\gaia, \gbp} & \sout{20} & -- \\
& \gaia, \grp & 32 & 1.00 & & \gaia, \grp & 19 & 1.49  \\
& \sout{KMTNet SAAO, I} & \sout{549} & -- & & OGLE, I & 368 & 1.01 \\
& \textbf{KMTNet CTIO, I} & \textbf{820} & \textbf{1.81} & & MOA, I & 2068 & 1.71 \\
& \textbf{KMTNet SSO, I} & \textbf{736} & \textbf{4.22} & &  MOA, V & 277 & 1.57  \\
& MOA, I & 2068 & 1.83 & & & & \\
& MOA, V & 277 & 2.64 & & & & \\
\addlinespace \hline  \addlinespace
\multirow{9}{*}{078} & & & & \multirow{9}{*}{079} & \gaia, \gmag & 24 & 5.41 \\
&  &  &  & & \gaia, \gbp & 25 & 1.59 \\
& \textbf{\gaia, \gmag} & \textbf{39} & \textbf{1.00} & & \gaia, \grp & 24 & 1.47 \\
& \textbf{\gaia, \gbp} & \textbf{38} & \textbf{1.00} & & OGLE, I & 600 & 1.19 \\
& \textbf{\gaia, \grp} & \textbf{38} & \textbf{1.76} & & \textbf{KMTNet SAAO, I} & \textbf{2923} & \textbf{4.59} \\
& OGLE, I & 730 & 1.01 & &  \textbf{KMTNet CTIO, I} & \textbf{4327} & \textbf{2.01} \\
& & & & & \textbf{KMTNet SSO, I} & \textbf{2678} & \textbf{2.97} \\
& & &  & &  MOA, I & 4315 & 1.57 \\
& & & & & MOA, V & 334 & 3.00 \\
\addlinespace \hline  \addlinespace
\multirow{8}{*}{088} & \gaia, \gmag & 33 & 1.00 & \multirow{8}{*}{089} & & & \\
&  \sout{\gaia, \gbp} & \sout{32} & -- & &  \gaia, \gmag & 24 & 1.00\\
& \gaia, \grp & 33 & 1.00 & & \sout{\gaia, \gbp} & \sout{25} & -- \\
& OGLE, I & 682 & 1.01 & & \textbf{\gaia, \grp} & \textbf{23} & \textbf{1.00} \\
& \textbf{KMTNet SAAO, I} & \textbf{548} & \textbf{3.53} & & OGLE, I & 934 & 1.00 \\
& \textbf{KMTNet CTIO, I} & \textbf{837} & \textbf{3.36} & &  \textbf{KMTNet SAAO, I} & \textbf{242} & \textbf{2.88}\\
& \textbf{KMTNet SSO, I} & \textbf{748} & \textbf{3.45} & & \textbf{KMTNet CTIO, I} & \textbf{385} & \textbf{1.87} \\
& MOA, I & 7974 & 1.36 & & \textbf{KMTNet SSO, I} & \textbf{341} & \textbf{2.53}\\
& MOA, V & 281 & 1.42 & &  & &  \\\\
\multirow{4}{*}{097} & \gaia, \gmag & 54 & 1.00 & \multirow{4}{*}{103} & \gaia, \gmag & 66 & 1.00 \\
& \sout{\gaia, \gbp} & \sout{52} & -- & & \gaia, \gbp & 59 & 1.00 \\
& \gaia, \grp & 52 & 1.00 & & \gaia, \grp & 63 & 1.00 \\
& OGLE, I & 791 & 1.00 & & OGLE, I & 144 & 1.19 \\
\addlinespace \hline  \addlinespace
\multirow{7}{*}{118} &  &  & & \multirow{7}{*}{127} & \gaia, \gmag & 27 & 1.16 \\
& \gaia, \gmag & 22 & 1.00 & & \sout{\gaia, \gbp} & \sout{16} & -- \\
& \gaia, \gbp & 22 & 1.00 & & \sout{\gaia, \grp} & \sout{16} & -- \\
& \sout{\gaia, \grp} & \sout{23} & -- & &  OGLE, I & 2677 & 1.01 \\
& OGLE, I & 168 & 1.30 & & \textbf{KMTNet SAAO, I} & \textbf{288} & \textbf{3.41} \\
& ASASSN, V & 23 & 2.55& & \textbf{KMTNet CTIO, I}& \textbf{409} & \textbf{3.83} \\
& &  &  & & \textbf{KMTNet SSO, I}& \textbf{377} & \textbf{1.71} \\
\addlinespace \hline  \addlinespace
\multirow{4}{*}{142} & \gaia, \gmag & 29 & 1.00 & \multirow{4}{*}{143} & \gaia, \gmag & 60 & -- \\
& \sout{\gaia, \gbp} & \sout{27} & -- & & \sout{\gaia, \gbp} & \sout{58} & -- \\
& \gaia, \grp & 25 & 1.00 & & \gaia, \gmag & 60 & --  \\
& OGLE, I & 638 & 1.44 & & OGLE, I & 122 & -- \\
\addlinespace \hline  \addlinespace
\multirow{9}{*}{155} & \gaia, \gmag & 47 & 1.00 
 & \multirow{9}{*}{196} &  &  & \\
& \sout{\gaia, \gbp} & \sout{45} & -- & &  &  &  \\
& \gaia, \grp & 45 & 1.00 & & \gaia, \gmag & 31 & 1.00 \\
& OGLE, I & 329  & 1.00 & & \sout{\gaia, \gbp} & \sout{25} & -- \\
& \textbf{KMTNet SAAO, I} & \textbf{237} & \textbf{5.16} & & \gaia, \grp  & 31 & 1.00 \\
& \textbf{KMTNet CTIO, I} & \textbf{373} & \textbf{2.08} & & OGLE, I & 956 & 1.06 \\
& \textbf{KMTNet SSO, I} & \textbf{329} & \textbf{3.56} & & & & \\
& MOA, I & 1881 & 1.71 & &  &  & \\
& MOA, V & 237 & 1.22 & & & & \\
\addlinespace \hline  \addlinespace
\multirow{8}{*}{212} & \gaia, \gmag & 59 & 1.00 
 & \multirow{8}{*}{230} &  &  & \\
& \textbf{\gaia, \gbp} & \textbf{54} & \textbf{1.00} & &  &  &  \\
& \textbf{\gaia, \grp} & \textbf{55} & \textbf{1.25} & & &  & \\
& GSA, \gmag & 80 & 1.00 & & \gaia, \gmag & 28 & 1.00 \\
& OGLE, I & 2140  & 1.21 & & \sout{\gaia, \gbp} & \sout{24} & -- \\
& \textbf{KMTNet SAAO, I} & \textbf{694} & \textbf{5.16} & & \sout{\gaia, \grp}  & \sout{25} & -- \\
& \textbf{KMTNet CTIO, I} & \textbf{956} & \textbf{2.37} & & OGLE, I & 2838 & 1.20 \\
& \textbf{KMTNet SSO, I} & \textbf{855} & \textbf{1.97} & & & & \\
\addlinespace \hline  \addlinespace
\multirow{7}{*}{259} &  &  & & \multirow{7}{*}{270} & \gaia, \gmag & 24 & 1.00 \\
& \gaia, \gmag & 58 & 1.00 & & \sout{\gaia, \gbp} & \sout{9} & -- \\
& \sout{\gaia, \gbp} & \sout{50} & -- & & \sout{\gaia, \grp} & \sout{9} & -- \\
& \gaia, \grp & 50 & 1.00 & &  OGLE, I & 1140 & 1.00 \\
& GSA, \gmag & 164 & 1.00 & & \textbf{KMTNet SAAO, I} & \textbf{551} & \textbf{1.91} \\
& OGLE, I & 191 & 1.20 & & \textbf{KMTNet CTIO, I}& \textbf{820} & \textbf{1.27} \\
& &  &  & & \textbf{KMTNet SSO, I}& \textbf{759} & \textbf{1.45} \\
\addlinespace \hline  \addlinespace
\multirow{7}{*}{275} &  &  & & \multirow{7}{*}{284} & \gaia, \gmag & 42 & 1.00 \\
& \gaia, \gmag & 40 & 1.00 & & \sout{\gaia, \gbp} & \sout{34} & -- \\
& \sout{\gaia, \gbp} & \sout{38} & -- & & \sout{\gaia, \grp} & \sout{37} & -- \\
& \gaia, \grp & 39 & 1.00 & &  OGLE, I & 1095 & 1.00 \\
& OGLE, I & 131 & 1.24 & & \textbf{KMTNet SAAO, I} & \textbf{748} & \textbf{3.68} \\
& &  &  & &  \textbf{KMTNet CTIO, I} & \textbf{992} & \textbf{2.24} \\
& &  &  & & \textbf{KMTNet SSO, I}& \textbf{921} & \textbf{2.69} \\
\addlinespace \hline  \addlinespace
\multirow{7}{*}{326} & \gaia, \gmag & 39 & 1.00 & \multirow{7}{*}{331} & \gaia, \gmag & 42 & 1.00 \\
& \sout{\gaia, \gbp} & \sout{22} & -- & & \sout{\gaia, \gbp} & \sout{22} & -- \\
& \sout{\gaia, \grp} & \sout{33} & -- & & \gaia, \grp & 36 & 1.00 \\
& OGLE, I & 791 & 1.00 & & OGLE, I & 1776 & 1.06 \\
& \textbf{KMTNet SAAO, I} & \textbf{531} & \textbf{2.00} & & \textbf{KMTNet SAAO, I} & \textbf{527} & \textbf{1.91} \\
& \textbf{KMTNet CTIO, I} & \textbf{823} & \textbf{1.40} & & \textbf{KMTNet CTIO, I}& \textbf{822} & \textbf{1.20} \\
& \textbf{KMTNet SSO, I} & \textbf{731} & \textbf{1.34} & & \textbf{KMTNet SSO, I}& \textbf{755} & \textbf{1.44} \\
\addlinespace \hline  \addlinespace
\multirow{9}{*}{343} & \gaia, \gmag & 20 & 1.00 & \multirow{9}{*}{353} & \gaia, \gmag & 18 & 1.00 \\
& \sout{\gaia, \gbp} & \sout{10} & -- & & \sout{\gaia, \gbp} & \sout{12} & -- \\
& \sout{\gaia, \grp} & \sout{10} & -- & & \sout{\gaia, \grp} & \sout{12} & -- \\
& OGLE, I & 725 & 1.00 & & OGLE, I & 506 & 1.00 \\
& \textbf{KMTNet SAAO, I} & \textbf{679} & \textbf{2.50} & & \textbf{KMTNet SAAO, I} & \textbf{657} & \textbf{2.22} \\
& \textbf{KMTNet CTIO, I} & \textbf{919} & \textbf{1.81} & & \sout{KMTNet CTIO, I}& \sout{241} & -- \\
& \textbf{KMTNet SSO, I} & \textbf{841} & \textbf{1.79} & & \textbf{KMTNet SSO, I}& \textbf{841} & \textbf{1.50} \\
& MOA, I & 10498 & 1.21 & & MOA, I & 5223 & 1.13\\
& MOA, V & 328 & 1.15 & & MOA, V & 191 & 1.14 \\
\addlinespace \hline  \addlinespace
\multirow{7}{*}{359} & \gaia, \gmag & 21 & 1.00 & \multirow{7}{*}{363} & \gaia, \gmag & 23 & 1.00 \\
& \sout{\gaia, \gbp} & \sout{17} & -- & & \sout{\gaia, \gbp} & \sout{19} & -- \\
& \gaia, \grp & 19 & 1.00 & & \sout{\gaia, \grp} & \sout{22} & -- \\
& OGLE, I & 553 & 1.00 & & OGLE, I & 824 & 1.00 \\
& \textbf{KMTNet SAAO, I} & \textbf{260} & \textbf{2.79} & & \textbf{KMTNet SAAO, I} & \textbf{780} & \textbf{1.68} \\
& \textbf{KMTNet CTIO, I} & \textbf{994} & \textbf{1.58} & & \textbf{KMTNet CTIO, I}& \textbf{386} & \textbf{1.39} \\
& \textbf{KMTNet SSO, I} & \textbf{895} & \textbf{1.88} & & \textbf{KMTNet SSO, I}& \textbf{364} & \textbf{1.45} \\
\end{longtable}
\end{ThreePartTable}

\clearpage
\LTcapwidth=\textwidth
\begin{ThreePartTable}
\begin{TableNotes}
\item{\textbf{Notes:} The columns are event name, mass function MF used as a prior for lens mass, the mass of the lens $M_\mathrm{L}$ in solar masses, distance to the lens $D_\mathrm{L}$ in kpc, the brightness of an MS star with $M_\mathrm{L}$ mass at a distance $D_\mathrm{L}$, the brightness of the blend $G_\mathrm{blend}$ obtained from the microlensing model, Einstein radius $\theta_\mathrm{E}$, maximal astrometric displacement $\delta_\mathrm{max}$, proper motion of the lens $\mu_\mathrm{L}$, transverse velocity $v_\mathrm{t}$, and probability for a dark lens. In the MF column, K2001 refers to \cite{2001Kroupa}, and M2021 to \cite{2021MrozBHOGLE}.}
\end{TableNotes}

\LTcapwidth=\textwidth
\begin{ThreePartTable}
\begin{TableNotes}
\item{\textbf{Notes:} The table contains the event name, its equatorial coordinates (RA, dec), proper motion in Right Ascension $\mu_\mathrm{RA}$ and declination $\mu_\mathrm{dec}$ from \gdr{3} and their correlation corr($\mu_\mathrm{RA}$, $\mu_\mathrm{dec}$), as well as the Renormalized Unit Weight Error (RUWE) (if available), and angular stellar radius $\theta_*$ found for the event solution. Column "Flag" marks if the event's colour is within the applicable range for the relation form \cite{2018Adams}. Column "Type" marks the assumed source star type for further analysis.}
\end{TableNotes}
\begin{longtable}{c c c c c c c c c c}
\caption{\label{tab:source_prop} %\changed{Values of $\theta_*$ uncertainties were changed.} 
Source star properties. The description of the columns is at the end of the table.}\\
\hline
\addlinespace
{\gaia}DR3- & RA & dec & $\mu_\mathrm{RA}$ & $\mu_\mathrm{Dec}$ & \multirow{2}{*}{corr($\mu_\mathrm{RA}$, $\mu_\mathrm{dec}$)} & \multirow{2}{*}{RUWE} & $\theta_*$ &  \multirow{2}{*}{Flag} & \multirow{2}{*}{Type} \\
ULENS- & [deg] & [deg] & [$\mathrm{mas}\,\mathrm{yr}^{-1}$] &  [$\mathrm{mas}\,\mathrm{yr}^{-1}$] & & & [$\mu \mathrm{mas}$] & & \\
\addlinespace
\hline
\hline
\addlinespace
\endfirsthead
\caption{continued.}\\
\hline
\addlinespace
{\gaia}DR3- & RA & dec & $\mu_\mathrm{RA}$ & $\mu_\mathrm{Dec}$ & \multirow{2}{*}{corr($\mu_\mathrm{RA}$, $\mu_\mathrm{dec}$)} & \multirow{2}{*}{RUWE} & $\theta_*$ &  \multirow{2}{*}{Flag} & \multirow{2}{*}{Type} \\
ULENS- & [deg] & [deg] & [$\mathrm{mas}\,\mathrm{yr}^{-1}$] &  [$\mathrm{mas}\,\mathrm{yr}^{-1}$] & & & [$\mu \mathrm{mas}$] & & \\
\addlinespace
\hline
\hline
\addlinespace
\endhead
\addlinespace
\hline
\endfoot
\addlinespace
\hline
\insertTableNotes
\endlastfoot
 025-G1- & 260.8781 & -27.3788 & $-4.61 \pm 0.06$ & $-5.56 \pm 0.04$ & $0.21$ & $1.08$ & $5.58 \pm 1.22$ & true & giant \\
 025-G2- & 260.8781 & -27.3788 & $-4.61 \pm 0.06$ & $-5.56 \pm 0.04$ & $0.21$ & $1.08$ & $6.22 \pm 1.44$ & true & giant \\
 035-G+ & 267.9705 & -34.9749 & $-4.37 \pm 0.11$ & $-3.95 \pm 0.07$ & $0.23$ & $1.83$ & $6.49 \pm 1.12$ & true & giant \\
 035-G- & 267.9705 & -34.9749 & $-4.37 \pm 0.11$ & $-3.95 \pm 0.07$ & $0.23$ & $1.83$ & $6.46 \pm 1.11$ & true & giant \\
 069-G+ & 269.0712 & -32.4500 & $-2.94 \pm 0.14$ & $-3.11 \pm 0.10$ & $0.41$ & $1.33$ & $6.19 \pm 1.03$ & true & giant \\
 069-G- & 269.0712 & -32.4500 & $-2.94 \pm 0.14$ & $-3.11 \pm 0.10$ & $0.41$ & $1.33$ & $6.34 \pm 1.05$ & true & giant \\
 073-G+ & 272.4690 & -30.0406 & $1.18 \pm 0.11$ & $-7.59 \pm 0.08$ & $0.30$ & $1.05$ & $3.47 \pm 0.83$ & true & giant \\
 073-G- & 272.4690 & -30.0406 & $1.18 \pm 0.11$ & $-7.59 \pm 0.08$ & $0.30$ & $1.05$ & $3.67 \pm 0.78$ & true & giant \\
\end{longtable}
\end{ThreePartTable}

\begin{longtable}{c c c c c c c c c c c}
\caption{\label{tab:lens_mass} Table of the lens mass and distance estimates of the 14~candidate dark lens microlensing events. The description of the columns is at the end of the table.}\\
% A machine-readable version of this table is available at the \kkcom{CDS}. \\
\hline \addlinespace
% {\gaia}DR3- & $A_G$ & \multirow{2}{*}{MF} & $D_\mathrm{L}$ & $M_\mathrm{L}$ & $G_\mathrm{MS, 0}$ & $G_\mathrm{MS, A}$ &  $G_\mathrm{blend}$ & $P_A$ & $P_0$\\
% ULENS & [mag] &  & [kpc] & [$\Msun$] & [mag] & [mag] &  [mag] & [\%] & [\%]\\
{\gaia}DR3- & \multirow{2}{*}{MF} & $M_\mathrm{L}$ & $D_\mathrm{L}$ & $G_\mathrm{MS}$ & $G_\mathrm{blend}$ & $\theta_\mathrm{E}$ & $\delta_\mathrm{max}$ & $\mu_\mathrm{L}$ & $v_\mathrm{t}$ & Prob \\
ULENS &  & [$\Msun$] & [kpc] & [mag] & [mag] & [mas] & [mas] & [$\mathrm{mas}\,\mathrm{yr}^{-1}$] & [$\mathrm{km}\,\mathrm{s}^{-1}$] & [\%]\\
\hline \hline
\noalign{\smallskip}
\endfirsthead
\caption{continued.}\\
\hline  \addlinespace
% {\gaia}DR3- & $A_G$ & \multirow{2}{*}{MF} & $D_\mathrm{L}$ & $M_\mathrm{L}$ & $G_\mathrm{MS, 0}$ & $G_\mathrm{MS, A}$ &  $G_\mathrm{blend}$ & $P_A$& $P_0$\\
{\gaia}DR3- & \multirow{2}{*}{MF} & $M_\mathrm{L}$ & $D_\mathrm{L}$ & $G_\mathrm{MS}$ & $G_\mathrm{blend}$ & $\theta_\mathrm{E}$ & $\delta_\mathrm{max}$ & $\mu_\mathrm{L}$ & $v_\mathrm{t}$ & Prob \\
ULENS &  & [$\Msun$] & [kpc] & [mag] & [mag] & [mas] & [mas] & [$\mathrm{mas}\,\mathrm{yr}^{-1}$] & [$\mathrm{km}\,\mathrm{s}^{-1}$]  & [\%]\\
\hline \hline
\endhead
\hline
\endfoot
\addlinespace
\hline
\insertTableNotes
\endlastfoot
025-G1- & K2001 & $0.37^{+0.35}_{-0.19}$ & $0.48^{+0.31}_{-0.20}$ & $18.94^{+3.43}_{-4.61}$ & $20.26^{+4.74}_{-1.47}$ & $2.17^{+2.04}_{-1.11}$ & $0.77^{+1.49}_{-0.37}$ & $12.19^{+2.04}_{-1.12}$ & $27.55^{+22.36}_{-14.17}$ &$68.68$\\ 
025-G1- & M2021 & $0.91^{+0.43}_{-0.41}$ & $0.24^{+0.16}_{-0.07}$ & $12.56^{+4.65}_{-2.80}$ & $20.26^{+4.74}_{-1.47}$ & $5.16^{+2.42}_{-2.30}$ & $1.83^{+2.68}_{-1.01}$ & $20.94^{+2.42}_{-2.31}$ & $24.09^{+18.77}_{-9.99}$ &$95.27$\\ 
025-G1- & $\sim M^{-1}$ & $0.53^{+0.55}_{-0.30}$ & $0.37^{+0.30}_{-0.16}$ & $16.77^{+4.25}_{-5.61}$ & $20.26^{+4.74}_{-1.47}$ & $3.09^{+3.11}_{-1.71}$ & $1.09^{+2.19}_{-0.49}$ & $14.78^{+3.11}_{-1.71}$ & $25.59^{+26.33}_{-14.42}$ &$79.61$\\ 
 \addlinespace \hline  \addlinespace
025-G2- & K2001 & $0.58^{+0.57}_{-0.24}$ & $1.04^{+0.78}_{-0.50}$ & $19.13^{+3.45}_{-5.78}$ & $25.00^{+0.00}_{-2.95}$ & $1.84^{+1.78}_{-0.73}$ & $0.65^{+1.28}_{-0.39}$ & $2.70^{+1.78}_{-0.74}$ & $13.26^{+18.73}_{-10.08}$ &$91.84$\\ 
025-G2- & M2021 & $1.42^{+1.24}_{-0.66}$ & $0.47^{+0.44}_{-0.20}$ & $11.97^{+4.92}_{-3.80}$ & $25.00^{+0.00}_{-2.96}$ & $4.42^{+3.86}_{-2.07}$ & $1.56^{+2.93}_{-0.83}$ & $3.87^{+3.86}_{-2.07}$ & $8.62^{+16.67}_{-8.28}$ &$99.03$\\ 
025-G2- & $\sim M^{-1}$ & $1.41^{+1.24}_{-0.66}$ & $0.47^{+0.44}_{-0.20}$ & $12.01^{+5.00}_{-3.84}$ & $25.00^{+0.00}_{-2.96}$ & $4.39^{+3.86}_{-2.07}$ & $1.55^{+2.92}_{-0.82}$ & $3.81^{+3.86}_{-2.07}$ & $8.53^{+16.68}_{-8.27}$ &$99.06$\\ 
 \addlinespace \hline  \addlinespace
035-G+ & K2001 & $0.55^{+0.39}_{-0.23}$ & $0.81^{+0.27}_{-0.23}$ & $18.58^{+2.84}_{-4.06}$ & $25.00^{+0.00}_{-0.00}$ & $1.72^{+1.21}_{-0.73}$ & $0.61^{+1.04}_{-0.35}$ & $15.16^{+1.21}_{-0.74}$ & $58.35^{+23.82}_{-19.28}$ &$99.18$\\ 
035-G+ & M2021 & $0.95^{+0.32}_{-0.34}$ & $0.59^{+0.20}_{-0.12}$ & $14.43^{+3.58}_{-2.08}$ & $25.00^{+0.00}_{-0.00}$ & $2.94^{+1.03}_{-1.07}$ & $1.04^{+1.40}_{-0.66}$ & $22.84^{+1.04}_{-1.08}$ & $63.96^{+24.50}_{-15.70}$ &$99.98$\\
\hline
\end{longtable}
\end{ThreePartTable}

\begin{ThreePartTable}
\begin{TableNotes}
\item{\textbf{Notes:} The columns are event name, mass function MF used as a prior for lens mass, the mass of the lens $M_\mathrm{L}$ in solar masses, distance to the lens $D_\mathrm{L}$ in kpc, the brightness of an MS star with $M_\mathrm{L}$ mass at a distance $D_\mathrm{L}$, the brightness of the blend $G_\mathrm{blend}$ obtained from the microlensing model, Einstein radius $\theta_\mathrm{E}$, maximal astrometric displacement $\delta_\mathrm{max}$, proper motion of the lens $\mu_\mathrm{L}$, transverse velocity $v_\mathrm{t}$, and probability for a dark lens. In the MF column, K2001 refers to \cite{2001Kroupa}, and M2021 to \cite{2021MrozBHOGLE}.}
\end{TableNotes}

\begin{longtable}{c c c c c c c c c c c}
\caption{\label{tab:lens_mass_edge} %\changed{Values in this table were changed.} 
Table of the lens mass and distance estimates of the five candidate dark lens microlensing events, that didn't pass the $\piE$ criterion, but were chosen to be analysed further. The description of the columns is at the end of the table.
% A machine-readable version of this table is available at the \kkcom{CDS}. 
}\\
\hline \addlinespace
% {\gaia}DR3- & $A_G$ & \multirow{2}{*}{MF} & $D_\mathrm{L}$ & $M_\mathrm{L}$ & $G_\mathrm{MS, 0}$ & $G_\mathrm{MS, A}$ &  $G_\mathrm{blend}$ & $P_A$ & $P_0$\\
% ULENS & [mag] &  & [kpc] & [$\Msun$] & [mag] & [mag] &  [mag] & [\%] & [\%]\\
{\gaia}DR3- & \multirow{2}{*}{MF} & $M_\mathrm{L}$ & $D_\mathrm{L}$ & $G_\mathrm{MS}$ & $G_\mathrm{blend}$ & $\theta_\mathrm{E}$ & $\delta_\mathrm{max}$ & $\mu_\mathrm{L}$ & $v_\mathrm{t}$ & Prob \\
ULENS &  & [$\Msun$] & [kpc] & [mag] & [mag] & [mas] & [mas] & [$\mathrm{mas}\,\mathrm{yr}^{-1}$] & [$\mathrm{km}\,\mathrm{s}^{-1}$] & [\%]\\
\hline \hline
\noalign{\smallskip}
\endfirsthead
\caption{continued.}\\
\hline  \addlinespace
% {\gaia}DR3- & $A_G$ & \multirow{2}{*}{MF} & $D_\mathrm{L}$ & $M_\mathrm{L}$ & $G_\mathrm{MS, 0}$ & $G_\mathrm{MS, A}$ &  $G_\mathrm{blend}$ & $P_A$& $P_0$\\
{\gaia}DR3- & \multirow{2}{*}{MF} & $M_\mathrm{L}$ & $D_\mathrm{L}$ & $G_\mathrm{MS}$ & $G_\mathrm{blend}$ & $\theta_\mathrm{E}$ & $\delta_\mathrm{max}$ & $\mu_\mathrm{L}$ & $v_\mathrm{t}$ & Prob \\
ULENS &  & [$\Msun$] & [kpc] & [mag] & [mag] & [mas] & [mas] & [$\mathrm{mas}\,\mathrm{yr}^{-1}$] & [$\mathrm{km}\,\mathrm{s}^{-1}$]  & [\%]\\
\hline \hline
\endhead
\hline
\endfoot
\addlinespace
\hline
\insertTableNotes
\endlastfoot
103-G+ & K2001 & $0.93^{+0.74}_{-0.41}$ & $4.64^{+1.71}_{-1.41}$ & $20.34^{+3.66}_{-2.84}$ & $25.00^{+0.00}_{-4.72}$ & $0.84^{+0.34}_{-0.30}$ & $0.30^{+0.42}_{-0.19}$ & $2.52^{+0.37}_{-0.34}$ & $55.36^{+28.66}_{-24.42}$ &$83.63$\\
103-G+ & M2021 & $1.91^{+4.86}_{-0.86}$ & $4.81^{+1.31}_{-1.34}$ & $16.93^{+2.80}_{-2.22}$ & $25.00^{+0.00}_{-4.72}$ & $1.00^{+0.34}_{-0.28}$ & $0.36^{+0.48}_{-0.26}$ & $1.67^{+0.38}_{-0.32}$ & $38.07^{+19.06}_{-17.99}$ &$97.56$\\ 
103-G+ & $\sim M^{-1}$ & $1.58^{+2.38}_{-0.75}$ & $4.79^{+1.31}_{-1.44}$ & $17.87^{+3.19}_{-2.84}$ & $25.00^{+0.00}_{-4.72}$ & $0.98^{+0.35}_{-0.30}$ & $0.35^{+0.47}_{-0.24}$ & $1.81^{+0.38}_{-0.34}$ & $41.03^{+19.86}_{-19.96}$ &$94.48$\\ 
 \addlinespace \hline  \addlinespace
103-G- & K2001 & $0.91^{+0.73}_{-0.40}$ & $4.63^{+1.76}_{-1.48}$ & $20.45^{+3.56}_{-2.90}$ & $25.00^{+0.00}_{-4.97}$ & $0.84^{+0.35}_{-0.31}$ & $0.30^{+0.42}_{-0.19}$ & $2.55^{+0.38}_{-0.35}$ & $55.91^{+29.64}_{-25.42}$ &$83.12$\\ 
103-G- & M2021 & $1.97^{+5.61}_{-0.93}$ & $4.77^{+1.34}_{-1.31}$ & $16.79^{+2.95}_{-2.26}$ & $25.00^{+0.00}_{-4.97}$ & $1.02^{+0.34}_{-0.31}$ & $0.36^{+0.48}_{-0.25}$ & $1.56^{+0.38}_{-0.35}$ & $35.26^{+18.43}_{-17.51}$ &$97.52$\\ 
103-G- & $\sim M^{-1}$ & $1.62^{+3.14}_{-0.79}$ & $4.76^{+1.37}_{-1.39}$ & $17.77^{+3.28}_{-2.93}$ & $25.00^{+0.00}_{-4.97}$ & $0.99^{+0.33}_{-0.30}$ & $0.35^{+0.46}_{-0.24}$ & $1.81^{+0.36}_{-0.34}$ & $40.90^{+19.98}_{-19.69}$ &$94.43$\\ 
 \addlinespace \hline  \addlinespace
155-G1- & K2001 & $0.82^{+0.55}_{-0.36}$ & $0.93^{+0.47}_{-0.30}$ & $16.57^{+4.27}_{-3.72}$ & $20.02^{+0.16}_{-0.13}$ & $2.29^{+1.51}_{-1.02}$ & $0.81^{+1.35}_{-0.45}$ & $13.57^{+1.52}_{-1.04}$ & $59.67^{+36.70}_{-24.18}$ &$78.47$\\ 
155-G1- & M2021 & $1.26^{+0.45}_{-0.42}$ & $0.70^{+0.28}_{-0.15}$ & $13.39^{+3.00}_{-1.67}$ & $20.02^{+0.16}_{-0.13}$ & $3.48^{+1.08}_{-1.18}$ & $1.23^{+1.61}_{-0.81}$ & $19.71^{+1.09}_{-1.20}$ & $65.19^{+30.22}_{-18.16}$ &$97.41$\\ 
155-G1- & $\sim M^{-1}$ & $1.13^{+0.50}_{-0.48}$ & $0.75^{+0.38}_{-0.19}$ & $14.12^{+4.42}_{-2.21}$ & $20.02^{+0.16}_{-0.13}$ & $3.12^{+1.29}_{-1.33}$ & $1.10^{+1.56}_{-0.63}$ & $17.80^{+1.30}_{-1.34}$ & $63.65^{+37.01}_{-21.10}$ &$91.21$\\ 
 \addlinespace \hline  \addlinespace
155-G2- & K2001 & $1.40^{+0.68}_{-0.47}$ & $2.45^{+0.49}_{-0.47}$ & $16.44^{+2.26}_{-1.96}$ & $19.88^{+0.11}_{-0.10}$ & $1.46^{+0.51}_{-0.39}$ & $0.52^{+0.70}_{-0.38}$ & $7.69^{+0.54}_{-0.43}$ & $89.27^{+24.33}_{-22.24}$ &$92.29$\\ 
155-G2- & M2021 & $1.79^{+0.81}_{-0.56}$ & $2.38^{+0.51}_{-0.49}$ & $15.35^{+1.76}_{-1.66}$ & $19.88^{+0.11}_{-0.10}$ & $1.68^{+0.56}_{-0.41}$ & $0.60^{+0.79}_{-0.45}$ & $8.24^{+0.59}_{-0.45}$ & $93.00^{+26.70}_{-24.09}$ &$98.78$\\ 
155-G2- & $\sim M^{-1}$ & $1.73^{+0.80}_{-0.57}$ & $2.39^{+0.51}_{-0.48}$ & $15.52^{+1.87}_{-1.76}$ & $19.88^{+0.11}_{-0.10}$ & $1.65^{+0.56}_{-0.42}$ & $0.58^{+0.78}_{-0.44}$ & $8.13^{+0.59}_{-0.46}$ & $92.04^{+26.36}_{-23.86}$ &$97.54$\\ 
 \addlinespace \hline  \addlinespace
212-GSA+ & K2001 & $0.92^{+0.82}_{-0.42}$ & $8.23^{+0.20}_{-0.58}$ & $22.43^{+3.53}_{-2.98}$ & $25.00^{+0.00}_{-0.00}$ & $0.41^{+0.20}_{-0.13}$ & $0.15^{+0.22}_{-0.10}$ & $7.71^{+0.28}_{-0.24}$ & $300.67^{+18.28}_{-30.41}$ &$76.06$\\ 
212-GSA+ & M2021 & $2.31^{+2.21}_{-1.23}$ & $7.90^{+0.56}_{-1.27}$ & $17.94^{+3.49}_{-1.79}$ & $25.00^{+0.00}_{-0.00}$ & $0.69^{+0.47}_{-0.25}$ & $0.24^{+0.41}_{-0.16}$ & $7.92^{+0.51}_{-0.32}$ & $296.55^{+40.20}_{-59.69}$ &$98.07$\\ 
212-GSA+ & $\sim M^{-1}$ & $1.80^{+2.29}_{-0.93}$ & $8.09^{+0.34}_{-1.35}$ & $19.24^{+3.57}_{-2.85}$ & $25.00^{+0.00}_{-0.00}$ & $0.60^{+0.49}_{-0.22}$ & $0.21^{+0.39}_{-0.14}$ & $7.94^{+0.53}_{-0.29}$ & $304.49^{+33.31}_{-61.94}$ &$93.12$\\ 
 \addlinespace \hline  \addlinespace
212-GSA- & K2001 & $1.12^{+0.84}_{-0.50}$ & $8.25^{+0.24}_{-0.86}$ & $21.34^{+3.52}_{-2.70}$ & $25.00^{+0.00}_{-0.00}$ & $0.46^{+0.20}_{-0.15}$ & $0.16^{+0.23}_{-0.11}$ & $7.89^{+0.28}_{-0.25}$ & $308.67^{+19.88}_{-41.91}$ &$84.96$\\ 
212-GSA- & M2021 & $1.94^{+1.97}_{-0.81}$ & $8.06^{+0.45}_{-1.50}$ & $18.68^{+2.61}_{-2.23}$ & $25.00^{+0.00}_{-0.00}$ & $0.64^{+0.49}_{-0.20}$ & $0.23^{+0.40}_{-0.16}$ & $8.01^{+0.53}_{-0.28}$ & $305.82^{+37.09}_{-67.56}$ &$98.72$\\ 
212-GSA- & $\sim M^{-1}$ & $1.75^{+1.86}_{-0.80}$ & $8.14^{+0.34}_{-1.46}$ & $19.46^{+2.72}_{-2.79}$ & $25.00^{+0.00}_{-0.00}$ & $0.59^{+0.46}_{-0.19}$ & $0.21^{+0.37}_{-0.14}$ & $8.18^{+0.50}_{-0.28}$ & $315.64^{+32.55}_{-67.40}$ &$95.84$\\ 
 \addlinespace \hline  \addlinespace
259-GSA+ & K2001 & $1.50^{+2.38}_{-0.88}$ & $2.52^{+1.45}_{-1.09}$ & $16.58^{+5.58}_{-4.10}$ & $20.20^{+0.42}_{-0.30}$ & $1.23^{+1.27}_{-0.61}$ & $0.43^{+0.89}_{-0.22}$ & $3.66^{+1.31}_{-0.68}$ & $43.63^{+40.84}_{-26.97}$ &$75.25$\\
259-GSA+ & M2021 & $5.41^{+7.23}_{-3.23}$ & $1.63^{+1.23}_{-0.77}$ & $11.96^{+3.04}_{-2.25}$ & $20.20^{+0.42}_{-0.30}$ & $2.74^{+2.78}_{-1.37}$ & $0.97^{+1.95}_{-0.48}$ & $5.57^{+2.80}_{-1.40}$ & $42.91^{+54.07}_{-31.12}$ &$98.59$\\ 
259-GSA+ & $\sim M^{-1}$ & $4.61^{+6.41}_{-2.81}$ & $1.74^{+1.28}_{-0.83}$ & $12.35^{+3.50}_{-2.41}$ & $20.19^{+0.42}_{-0.30}$ & $2.47^{+2.62}_{-1.26}$ & $0.88^{+1.80}_{-0.43}$ & $5.13^{+2.63}_{-1.30}$ & $42.31^{+52.75}_{-30.93}$ &$96.37$\\ 
 \addlinespace \hline  \addlinespace
259-GSA- & K2001 & $1.05^{+1.73}_{-0.58}$ & $2.37^{+1.40}_{-1.04}$ & $18.05^{+5.38}_{-4.89}$ & $20.43^{+0.61}_{-0.37}$ & $1.14^{+1.20}_{-0.55}$ & $0.40^{+0.83}_{-0.21}$ & $4.00^{+1.23}_{-0.63}$ & $45.01^{+40.49}_{-26.84}$ &$66.45$\\ 
259-GSA- & M2021 & $4.26^{+6.37}_{-2.58}$ & $1.57^{+1.20}_{-0.75}$ & $12.23^{+3.50}_{-2.45}$ & $20.43^{+0.61}_{-0.37}$ & $2.61^{+2.61}_{-1.28}$ & $0.92^{+1.85}_{-0.47}$ & $6.06^{+2.63}_{-1.31}$ & $45.15^{+53.93}_{-31.28}$ &$97.76$\\ 
259-GSA- & $\sim M^{-1}$ & $3.52^{+5.54}_{-2.26}$ & $1.69^{+1.27}_{-0.82}$ & $12.76^{+4.44}_{-2.70}$ & $20.43^{+0.61}_{-0.37}$ & $2.33^{+2.45}_{-1.21}$ & $0.82^{+1.69}_{-0.39}$ & $5.60^{+2.47}_{-1.25}$ & $44.87^{+53.48}_{-31.72}$ &$93.48$\\ 
 \addlinespace \hline  \addlinespace
331-G+ & K2001 & $2.89^{+4.63}_{-1.52}$ & $6.49^{+1.19}_{-1.31}$ & $18.29^{+2.62}_{-1.11}$ & $25.00^{+0.00}_{-0.00}$ & $1.05^{+0.40}_{-0.33}$ & $0.37^{+0.51}_{-0.26}$ & $0.92^{+1.03}_{-1.00}$ & $28.44^{+36.88}_{-36.62}$ &$96.93$\\ 
331-G+ & M2021 & $14.01^{+45.92}_{-9.45}$ & $7.36^{+1.18}_{-1.13}$ & $17.43^{+0.18}_{-0.50}$ & $25.00^{+0.00}_{-0.00}$ & $1.47^{+0.54}_{-0.42}$ & $0.52^{+0.71}_{-0.37}$ & $1.80^{+1.09}_{-1.04}$ & $62.65^{+48.20}_{-45.80}$ &$99.96$\\ 
331-G+ & $\sim M^{-1}$ & $10.84^{+20.33}_{-7.33}$ & $7.18^{+1.08}_{-1.14}$ & $17.45^{+0.47}_{-0.53}$ & $25.00^{+0.00}_{-0.00}$ & $1.41^{+0.57}_{-0.40}$ & $0.50^{+0.70}_{-0.36}$ & $1.60^{+1.11}_{-1.03}$ & $54.38^{+45.86}_{-43.70}$ &$99.85$\\
 \addlinespace \hline  \addlinespace
331-G- & K2001 & $3.19^{+5.29}_{-1.73}$ & $6.62^{+1.16}_{-1.27}$ & $18.09^{+2.63}_{-0.93}$ & $25.00^{+0.00}_{-0.00}$ & $1.06^{+0.42}_{-0.34}$ & $0.38^{+0.53}_{-0.26}$ & $0.98^{+1.04}_{-1.01}$ & $30.65^{+37.92}_{-37.46}$ &$97.38$\\ 
331-G- & M2021 & $14.72^{+38.14}_{-9.71}$ & $7.45^{+1.17}_{-1.16}$ & $17.43^{+0.18}_{-0.50}$ & $25.00^{+0.00}_{-0.00}$ & $1.46^{+0.55}_{-0.42}$ & $0.52^{+0.71}_{-0.37}$ & $1.83^{+1.10}_{-1.04}$ & $64.75^{+49.00}_{-46.73}$ &$99.97$\\ 
331-G- & $\sim M^{-1}$ & $11.91^{+33.38}_{-7.93}$ & $7.36^{+1.17}_{-1.18}$ & $17.45^{+0.28}_{-0.50}$ & $25.00^{+0.00}_{-0.00}$ & $1.43^{+0.54}_{-0.41}$ & $0.51^{+0.70}_{-0.36}$ & $1.78^{+1.09}_{-1.03}$ & $62.04^{+47.97}_{-45.96}$ &$99.89$\\ 
\addlinespace
\hline
\end{longtable}
\end{ThreePartTable}

\LTcapwidth=\textwidth
\begin{ThreePartTable}
\begin{TableNotes}
\item{\textbf{Notes:} The table contains parameters for two types of microlensing point source-point lens models: with and without parallax. The non-parallax PSPL model parameters are: $t_\mathrm{0}$ -- the time of the peak of brightness, $u_\mathrm{0}$ -- corresponding separation of the lens and source at $t_\mathrm{0}$, $\tE$ -- Einstein timescale of the event, $I_{0, G}$ -- brightness in baseline in the \gmag-band, $f_{b, G}$ -- a fraction of the total flux at baseline belonging to the blend in the \gmag-band. The parallax model adds two additional parameters: $\piEN$ and $\piEE$, which are north and east components of the microlensing parallax vector. $t_\mathrm{0, par}$ is a non-fitted parameter, which defines the coordinate system for parallax measurement. Both $t_\mathrm{0}$ and $t_\mathrm{0, par}$ are in HJD' = HJD - 2450000. B is a string of letters denoting which type of \gaia data was used: "G" for events where we used the \gdr{3} photometry, and "GSA" for events where we opted for a \gaia Science Alerts lightcurve, finally C is a sign of the $u_\mathrm{0}$ of the solution ("+" for positive and "-" for negative). If there was more than one solution with the same $u_\mathrm{0}$ sign, we numbered them starting from one.}
\end{TableNotes}
\begin{longtable}{c c c c c c c c c c}
\caption{\label{tab:fit_best_model} %\changed{Values in this table were changed.} 
Parameters of all best-fitting solutions of the 35~analysed events. The description of the columns is at the end of the table.}\\
\hline \addlinespace
{\gaia}DR3- & \multirow{2}{*}{$t_\mathrm{0, par}$} & \multirow{2}{*}{$t_\mathrm{0}$} & \multirow{2}{*}{$u_\mathrm{0}$} & \multirow{2}{*}{$\tE$} & \multirow{2}{*}{$\piEN$} & \multirow{2}{*}{$\piEE$} & \multirow{2}{*}{$I_{0, G}$} & \multirow{2}{*}{$f_{b, G}$} & \multirow{2}{*}{$\chi^2$}\\
ULENS- & & & & & & & & & \\
\addlinespace
\hline \hline
\addlinespace
\endfirsthead
\caption{continued.}\\
\hline
\addlinespace
{\gaia}DR3- & \multirow{2}{*}{$t_\mathrm{0, par}$} & \multirow{2}{*}{$t_\mathrm{0}$} & \multirow{2}{*}{$u_\mathrm{0}$} & \multirow{2}{*}{$\tE$} & \multirow{2}{*}{$\piEN$} & \multirow{2}{*}{$\piEE$} & \multirow{2}{*}{$I_{0, G}$} & \multirow{2}{*}{$f_{b, G}$} & \multirow{2}{*}{$\chi^2$}\\
ULENS- & & & & & & & & & \\
\addlinespace
\hline \hline
\addlinespace
\endhead
\addlinespace
\hline
\endfoot
\addlinespace
\hline
\insertTableNotes
\endlastfoot
003-G0 &        -- &        $7297.87^{+0.10}_{-0.11}$ &        $0.05^{+0.01}_{-0.01}$ &        $60.60^{+5.49}_{-4.66}$ &        -- &        -- &        $14.16^{+0.01}_{-0.01}$ &        $0.18^{+0.10}_{-0.10}$ &        263.76\\
003-G1+ &        7297.00 &        $7298.59^{+0.31}_{-0.31}$ &        $0.04^{+0.01}_{-0.01}$ &        $63.19^{+12.48}_{-7.14}$ &        $-0.57^{+0.34}_{-0.24}$ &        $0.19^{+0.07}_{-0.06}$ &        $14.16^{+0.01}_{-0.01}$ &        $0.23^{+0.17}_{-0.15}$ &        255.64 \\ 

003-G2+ &        7297.00 &        $7298.63^{+0.33}_{-0.31}$ &        $0.05^{+0.01}_{-0.01}$ &        $63.79^{+15.87}_{-7.11}$ &        $0.56^{+0.25}_{-0.35}$ &        $0.09^{+0.07}_{-0.07}$ &        $14.16^{+0.01}_{-0.01}$ &        $0.21^{+0.19}_{-0.14}$ &        267.84 \\ 

003-G1- &        7297.00 &        $7298.68^{+0.35}_{-0.31}$ &        $-0.05^{+0.01}_{-0.01}$ &        $63.06^{+13.96}_{-6.50}$ &        $-0.57^{+0.37}_{-0.26}$ &        $0.19^{+0.07}_{-0.06}$ &        $14.16^{+0.01}_{-0.01}$ &        $0.19^{+0.18}_{-0.13}$ &        266.68 \\ 

003-G2- &        7297.00 &        $7298.50^{+0.32}_{-0.29}$ &        $-0.04^{+0.01}_{-0.01}$ &        $64.02^{+15.38}_{-7.74}$ &        $0.54^{+0.24}_{-0.32}$ &        $0.08^{+0.07}_{-0.07}$ &        $14.16^{+0.01}_{-0.01}$ &        $0.24^{+0.19}_{-0.16}$ &        256.42 \\ 

018-G0 &        -- &        $7721.31^{+0.20}_{-0.21}$ &        $0.93^{+0.03}_{-0.04}$ &        $79.88^{+2.21}_{-1.50}$ &        -- &        -- &        $15.99^{+0.01}_{-0.01}$ &        $0.06^{+0.08}_{-0.06}$ &        21375.63\\
018-G+ &        7721.00 &        $7712.63^{+1.40}_{-1.37}$ &        $0.19^{+0.02}_{-0.02}$ &        $321.86^{+27.75}_{-29.24}$ &        $-0.06^{+0.01}_{-0.01}$ &        $-0.10^{+0.01}_{-0.01}$ &        $16.00^{+0.01}_{-0.01}$ &        $0.92^{+0.01}_{-0.01}$ &        20962.81 \\
\end{longtable}
\end{ThreePartTable}
\twocolumn

\end{appendix}

\end{document}